\documentclass[preprint,a4paper,5p,times,twocolumn,sort&compress]{elsarticle}

\usepackage{amssymb}
\usepackage[utf8]{inputenc}
\usepackage{graphicx}
\usepackage{hyperref}
\usepackage{float}
\usepackage{amsmath}
\usepackage{amssymb}
\usepackage[T1]{fontenc}
\usepackage{hyperref} %% for \autoref
\usepackage{isotope} %% for \isotope
\usepackage{upgreek}
\usepackage{sidecap}
\usepackage{wrapfig}
\usepackage{placeins}
\usepackage{subcaption}
\usepackage{booktabs}
\usepackage{color}

\usepackage{lineno}

\usepackage{xspace}
\def\cd113{\isotope[113][]{Cd}}
\def\gA{$g_\text{A}$\xspace}

\journal{Physics Letters B}

\begin{document}

\begin{frontmatter}



\title{Confirmation of $g_{\rm A}$ quenching using the revised spectrum-shape method for the analysis of the $^{113}$Cd $\beta$-decay as measured with the COBRA demonstrator}


\author[jyvaskyla]{Joel~Kostensalo}
\author[jyvaskyla]{Jouni~Suhonen}
\author[dresden]{Juliane~Volkmer}
\author[dresden]{Stefan~Zatschler\corref{corauthor}\fnref{toronto}}
\cortext[corauthor]{corresponding author}
\ead{stefan.zatschler@utoronto.ca}
\author[dresden]{Kai~Zuber}

\address[jyvaskyla]{University of Jyvaskyla, Department of Physics, P.O. Box 35, FI-40014, Finland}
\address[dresden]{TU Dresden, Institute of Nuclear and Particle Physics, Zellescher Weg 19, 01069 Dresden, Germany}
\fntext[toronto]{Present affiliation: Department of Physics, University of Toronto, Toronto, ON, M5S 1A7, Canada}

% PAGE LIMIT PLB
%The total length of the paper should preferably not exceed six journal pages, excluding the list of authors, abstract, references, figure captions and three figures. In the case that more figures are required, the text should be shortened accordingly.

\begin{abstract}
In this article we present an updated spectrum-shape analysis of the $^{113}$Cd fourfold forbidden non-unique $\beta$-decay transition in order to address the quenching of the weak axial-vector coupling $g_{\rm A}$ in low-momentum exchange nuclear processes.
The experimental data were collected in a dedicated low-threshold run with the COBRA demonstrator at the LNGS and resulted in 44 individual $^{113}$Cd spectra.
These data are evaluated in the context of three nuclear model frameworks based on a revised version of the spectrum-shape method and the conserved vector current hypothesis.
The novel idea devised in the present work is to fit the value of the small relativistic nuclear matrix element (s-NME) driving the nuclear model calculations, which remained essentially as a free parameter in previous studies.
This is done by tuning the nuclear structure calculations and making use of the interplay of $g_{\rm A}$ and the s-NME such that the experimentally known $^{113}$Cd half-life gets reproducible by the different frameworks.
In this way, a best fit s-NME value can be derived for each of the considered nuclear models, which finally enters the template calculations used to perform the spectrum-shape analysis for each of the obtained $^{113}$Cd spectra.
The primary analysis strategy results in significantly quenched values of the axial-vector coupling for all three nuclear models: $\overline{g}_{\rm A}(\text{ISM}) = 0.907 \pm 0.064$, $\overline{g}_{\rm A}(\text{MQPM}) = 0.993 \pm 0.063$ and $\overline{g}_{\rm A}(\text{IBFM-2}) = 0.828 \pm 0.140$.
Moreover, with our data-driven approach one of the main shortcomings of the spectrum-shape method has been resolved.
This achievement is a milestone in the description of strongly forbidden $\beta$-decays and adds to the indications for the existence of a quenching of $g_{\rm A}$ in low-momentum exchange nuclear processes.
\end{abstract}


\end{frontmatter}

%\linenumbers

\section{Introduction}
The discovery of neutrino masses through the observation of flavor oscillations boosted the importance of direct and indirect neutrino mass searches -- including the search for the hypothesized neutrinoless double beta decay ($0\nu\beta\beta$-decay).
Nowadays, the hunt for this ultra-rare nuclear transition is one of the most active research fields at the intersection of nuclear, particle and astroparticle physics.
Its main and most evident feature is the explicit violation of the total lepton number -- an accidentally conserved quantity in the Standard Model of particle physics -- by two units.
Furthermore, its observation would prove the Majorana nature of neutrinos \cite{Schechter_Valle_1982, Duerr2011}.
This, in turn, would support the exciting theoretical explanation for the origin of the observed baryon asymmetry in the universe through the process of leptogenesis \cite{Primakoff1981} and, potentially, shed light on the role of neutrinos in the early universe's structure formation.
For these reasons, $0\nu\beta\beta$-decay searches provide a key tool for probing the fundamental nature of neutrinos and could add to the indications for physics beyond the Standard Model.

The predicted rate for the $0\nu\beta\beta$-decay, in particular if mediated by the exchange of a light Majorana neutrino, depends strongly on the numerical value of the weak axial-vector coupling \gA, appearing in the leading Gamow-Teller (GT) part of the nuclear matrix element (NME), denoted as $\mathcal{M}^{0\nu}$.
The relation can be expressed as follows, using the involved NME, a kinematic phase space factor $\mathcal{G}^{0\nu}$ -- which is well known, the effective Majorana neutrino mass $m_{\beta\beta}$ and the electron mass $m_e$:
\begin{eqnarray}
\left( T_{1/2}^{0\nu} \right)^{-1} &=& g_\text{A}^4 \cdot \mathcal{G}^{0\nu}(Q_{\beta\beta}, Z) \cdot \left\vert \mathcal{M}^{0\nu}(A,Z) \right\vert^2 \cdot \left\vert \frac{m_{\beta\beta}}{m_e} \right\vert^2 \\
\text{with }\mathcal{M}^{0\nu} &=& \mathcal{M}^{0\nu}_\text{GT} - \left( \frac{g_\text{V}}{g_\text{A}} \right)^2 \cdot \mathcal{M}^{0\nu}_\text{F} - \mathcal{M}^{0\nu}_\text{T}.
\end{eqnarray}

In general, such a strong \gA dependence does not only apply to $0\nu\beta\beta$-decays, but also to $2\nu\beta\beta$-decays and GT-dominated single $\beta$-decays.
For those processes, however, the involved Fermi~($\mathcal{M}_\text{F}$) and tensor parts~($\mathcal{M}_\text{T}$) are usually sub-leading.

A wide set of nuclear theory frameworks has been adopted to calculate the corresponding NMEs (see the review in \cite{Suhonen_report_2019}).
Both in the $\beta$-decay and the $2\nu\beta\beta$-decay calculations the GT matrix elements are consistently found to be too large to reproduce the experimental half-lives \cite{Suhonen_gA_review2017}.
This issue can be ``\textit{cured}'' by reducing the strength of the axial-vector coupling $g_{\rm A}$ \cite{Suhonen_gA_review2017}.
One of the common approaches is to introduce an effective \gA in place of the free value $g_\text{A}^\text{free} = 1.276(4)$ \cite{UCNA2010}.
This phenomenological modification is referred to as the \textit{quenching} or \textit{renormalization} of \gA.
However, in the calculation of NMEs the associated quenching of \gA is usually not addressed in a quantitative way.

Only quite recently the subject was discussed in the scope of \textit{ab initio} nuclear many-body frameworks \cite{Hergert2020}.
In these theories quite promising results have been obtained towards solving the quenching problem for very light nuclei ($A \le 10$) in the framework of the Green's function Monte Carlo approach \cite{Pastore2018, King2020} with consistently constructed
local chiral interactions and currents.
For the light and very light nuclei ($A < 50$) the quenching problem seems to be solved by shifting correlations from the wave functions into induced higher-order contributions to the renormalized transition operator (adding the two-body chiral currents) in the no-core shell model and the valence space in-medium similarity renormalization group approaches \cite{Gysbers2019}.
Further work towards medium-heavy and heavy open-shell nuclei is needed in order to
tackle the plenitude of further interesting cases for the $g_{\rm A}$ quenching \cite{Suhonen_gA_review2017}.

Additionally, systematic studies of the \gA quenching were addressed in low-momentum exchange processes such as single $\beta$-decays and $2\nu\beta\beta$-decays (see e.g.~Ref.~\cite{Suhonen_gA_review2017}).
As the $0\nu\beta\beta$-decay is a high-momentum exchange process involving states up to $\sim 100$\,MeV, it is not clear how the results obtained for the quenching of \gA in the low-momentum exchange processes can be translated to the $0\nu\beta\beta$ case.
Despite this unknown correlation, it is exceptionally important to study the quenching of \gA in as many ways as possible to identify its origin and to help clarifying its impact on new physics phenomena such as the potential observation of $0\nu\beta\beta$-decay.

Different methods were proposed to quantify the quenching effect in decay processes with low-momentum exchange such as discussed in the review articles \cite{Suhonen_gA_review2017} and \cite{Engel2017}.
At low energies, the quenching of \gA has several potential sources, including non-nucleonic degrees of freedom (e.g.~delta resonances) and giant multipole resonances (like the GT giant resonance).
Both reduce the transition strength for the lowest excited nuclear states.
Further sources of quenching (or sometimes also enhancement, see Ref.~\cite{Kostensalo2018}) are nuclear processes beyond the impulse approximation (in-medium meson-exchange or two-body weak currents) and deficiencies in the handling of the nuclear many-body problem (too small single-particle valence spaces, lacking many-body configurations, omission of three-body nucleon-nucleon interactions, etc.).

One of the proposed methods exploits the dependence of the $\beta$-spectrum shape of highly forbidden non-unique decays on~\gA.
First direct evidence for a quenching of \gA in this regard has been demonstrated in \cite{Cobra_Cd113_2020} by analyzing the low-energy \cd113 background data of the \mbox{COBRA} demonstrator.
The present article describes an updated interpretation of the experimental data in the context of improved nuclear model calculations.

\section{The spectrum-shape method}
\label{sec:SSM}

\subsection{General aspects}
In our previous article on the electron spectral shapes of the $\beta$-decay of
$^{113}$Cd \cite{Cobra_Cd113_2020}, the formalism of the spectrum-shape method (SSM) \cite{Haaranen2016} was used to access the effective value of the axial-vector coupling $g_{\rm A}$.
In the same reference an extensive account of the SSM formalism was given, and hence we are going to give only a brief outline of the main philosophy of the approach in the following.

The SSM is based on the fact that for forbidden non-unique $\beta$-decays the energy spectrum of the emitted electrons -- described by the $\beta$-decay shape function -- depends on the leptonic phase-space factors and the nuclear matrix elements (NMEs) in a complex way.
The phase-space factors can be calculated to a desired accuracy \cite{Mustonen2006}, but the NMEs are subject to systematic uncertainties of the underlying nuclear models.

The complexities of the shape function are condensed in the so-called shape factor $C(w_e)$, which is a function of the total energy $w_e$ of the emitted electron (in units of the electron rest mass) \cite{Suhonen_gA_review2017, Haaranen2016}.
The shape factor can be divided into an axial-vector part $C_{\rm A}(w_e)$, a vector part $C_{\rm V}(w_e)$ and a mixed vector-axial-vector part $C_{\rm VA}(w_e)$, such that we obtain
\begin{equation}
C(w_e) = g_{\rm A}^2\left[C_{\rm A}(w_e) + \left(\frac{g_{\rm V}}{g_{\rm A}}\right)^2C_{\rm V}(w_e) + \frac{g_{\rm V}}{g_{\rm A}}C_{\rm VA}(w_e) \right] \,.
\label{eqn:shape_factor}
\end{equation} 

The interference between the sum of the $C_{\rm A}$ and $C_{\rm V}$ part with the mixed $C_{\rm VA}$ part cause a variation of the electron spectral shapes depending on the ratio $g_{\rm V}/g_{\rm A}$.
In \cite{Haaranen2016} it was noticed that the variation was rather strong for the $\beta$-decay transition $^{113}\textrm{Cd}(1/2^+) \to \,^{113}\textrm{In}(9/2^+)$ when the NMEs were computed by using the interacting shell model (ISM) and the microscopic quasiparticle-phonon model (MQPM) \cite{Toivanen1998}.
%Moreover, the dependence was found to be very sensitive to the value of $g_{\rm A}$.
In a further study of this transition \cite{Haaranen2017}, a similar dependence was recorded for the microscopic interacting boson-fermion model (IBFM-2). 

The essence of the SSM philosophy is to compare the electron spectral shapes computed in dependence on \gA with the measured one in order to be able to access an effective value of $g_{\rm A}$.
In \cite{Haaranen2017} the computed electron spectra could be compared with the one measured in \cite{Belli2007}.
It was found that the measured spectrum was roughly reproduced by the calculated spectra of all three nuclear models for values of $g_{\rm A}/g_{\rm V}\sim 0.9$.
This is remarkable considering the totally different nuclear-structure principles behind these models.
Following this, the result could be verified in our previous COBRA study \cite{Cobra_Cd113_2020} with a full evaluation of the experimental systematic uncertainties.
Further extensive studies have been carried out in order to find similar cases of a high sensitivity to the $g_{\rm A}/g_{\rm V}$ ratio \cite{Kostensalo2018, Kostensalo2017b, Suhonen2017c, Suhonen2019, Kirsebom2019, Kumar2020}.

\subsection{CVC-inspired improvement of the SSM}
\label{sec:CVC_improvement}
A particular problem with the earlier SSM calculations \cite{Cobra_Cd113_2020, Haaranen2016, Haaranen2017} was that the nuclear models could not reproduce the experimental half-life of $^{113}$Cd and the measured electron spectral shape at the same time.
It was speculated that one source of this discrepancy could be the computed NMEs, which are prone to the inaccuracies of the adopted nuclear models.
In \cite{Kirsebom2019} similar inaccuracies were recorded in the calculation of the branching ratios and electron spectral shape for the second-forbidden non-unique $\beta^-$ transition $^{20}\textrm{F}(2^+)\to\,^{20}\textrm{Ne}(0^+)$.

Of particular interest for the SSM are the so-called small relativistic NME (s-NME) and the large vector-type NME (l-NME).
As pointed out in \cite{Behrens1982}, the magnitude of the s-NME, denoted as $^{V}\mathcal{M}^{(0)}_{K K-1 1}$, can be related to the l-NME, denoted as $^{V}\mathcal{M}^{(0)}_{KK0}$, by the conserved vector current (CVC) hypothesis:
\begin{eqnarray}
^V\mathcal{M}^{(0)}_{K K-1 1} &= \frac{1}{\sqrt{K(2K+1)}}\left\lbrack
\frac{(w_e -\Delta m_{np})R}{\hbar c} + \frac{6}{5}\alpha Z\right\rbrack\,
^V\mathcal{M}^{(0)}_{K K 0}/R .
\label{eqn:sNME}
\end{eqnarray}
Here $w_e=m_e + Q_\beta = 0.833$\,MeV is the total energy of a \mbox{$\beta$-decay} electron, $R=1.2\cdot A^{1/3}$\,fm is the nuclear radius and $Z$ the nuclear charge, \mbox{$\hbar c = 197.3\,\textrm{ MeV\, fm}$}, $\alpha=1/137$, and \mbox{$\Delta m_{np} = M_n-M_p = 1.293$\,MeV} is the mass difference between a neutron and a proton.
For the \cd113 decay transition we have $K=4$ and follow the definitions of the NMEs as presented in \cite{Behrens1982}.
This results in l-NME $^{V}\mathcal{M}^{(0)}_{440}$ given in units of fm$^4$ whereas the s-NME $^{V}\mathcal{M}^{(0)}_{431}$ is given in units of fm$^3$.
The $Z$ term is an approximation to the Coulomb displacement energy and the whole expression inside the square brackets is an approximation for the excitation energy
of the isobaric analog state.

The relation in Eqn.~(\ref{eqn:sNME}) is exact for ``ideal'' NMEs, but becomes only approximate for NMEs computed under the assumption of the impulse approximation and using approximate nuclear model wave functions.
Because many or all of the non-zero contributions come from outside the canonical shell-model valence space, it is hard to calculate the s-NME accurately. 
The l-NME, on the other hand, is easier to evaluate since it is usually not subject to this problem.
Despite its smallness, the value of the s-NME can strongly affect the calculated half-life and spectral shape of a forbidden non-unique transition \cite{Kirsebom2019, Kumar2020, Behrens1982}.

In some ISM and IBFM-2 calculations, the s-NME is found to be exactly zero due to the limitations of the single-particle model space.
This is also the case in the present calculations before applying our data-driven approach.
On the other hand, in MQPM calculations the same s-NME is typically non-zero; in our calculations it is on the order of 0.4.
Moreover, in the MQPM it is necessary to adjust the single-particle energies using experimental data, which can only be done for the orbitals close to the Fermi surface.
Thus, the contributions which are outside the shell-model space are accounted for to some degree, but for individual states this may not be very accurate.

At this point it is worth pointing out that the use of Eqn.~(\ref{eqn:sNME}) can be dangerous due to model
dependence of the expression inside the square brackets, as discussed critically in \cite{Damgaard1966}. It is particularly dangerous in the present case since the first term inside the square brackets is negative and thus interferes destructively with the second one.
Owing to the above-mentioned dangers of using the exact s-NME values predicted by Eqn.~(\ref{eqn:sNME}) we, instead, adopt the philosophy of Kumar \textit{et al.} \cite{Kumar2020} and treat the s-NME as a free parameter to be fit to the experimental half-life of \cd113.
This is referred to as the \textit{half-life method} in the following.
See section~\ref{sec:sNME_fit} for a description of the exact procedure.
The resulting CVC-inspired values are about $\text{s-NME} \sim 0.2-0.6$ depending on the nuclear models used in this study.
The range of values emerges from the use of the different values of $g_{\rm A}$ in the spectrum-shape analyses and the associated separate fixing of the half-life for each value of $g_{\rm A}$ and nuclear model.

While this approach is able to reduce the observed tension between the spectrum-shape and half-life method with regards to a quenching of \gA, it does not fully resolve them.
This is why the original procedure has been further improved in the course of the present work.
By combining the results of the spectrum-shape analysis and the correlation of the s-NME and \gA with respect to the \cd113 decay's half-life, it is possible to derive a best fit s-NME value for each nuclear model (see section~\ref{sec:sNME_fit}).
In this way, we have created an ``enhanced'' SSM, and with this tool we can now attack the quenching problem of $g_{\rm A}$ in a more consistent way.
Furthermore, since the present SSM is ``CVC-inspired'', we keep the CVC-compatible value $g_{\rm V}=1.00$ of the vector coupling strength in all our calculations.
An illustration of the predicted \gA dependence of the \cd113 spectrum-shape within the revised SSM under the CVC hypothesis for the three considered nuclear models (ISM, MQPM, IBFM-2) can be found in Fig.~\ref{fig:appendix_gA_dependence_cvc} of the appendix.

\section{The COBRA experiment}
The COBRA collaboration uses room temperature CdZnTe (CZT) semiconductor detectors \cite{Zuber2001} to search for rare single $\beta$-decays and $\beta\beta$-decays.
The detector material contains several isotopes of interest due to its natural composition.
Because of the typical half-lives for such rare nuclear processes being in the order of the age of the universe and beyond, the detector setup needs to be well-shielded from backgrounds that could mimic the signal processes.
For this reason COBRA is located at the Italian Laboratori Nazionali del Gran Sasso (LNGS) which is shielded against cosmic rays by an average rock overburden of 1400\,m.
From Sept.'11 to Nov.'19, an array of 64 coplanar-grid (CPG) CZT detectors has been operated at the LNGS.
This array is referred to as the \mbox{COBRA} demonstrator \cite{Cobra_demonstrator_2016}.
Each of the CZT crystals has a size of about 1$\times$1$\times$1\,cm$^3$ and a mass of about $6.0\,$g.
Based on the knowledge gathered during the demonstrator's operation, the experiment was upgraded to the extended demonstrator COBRA XDEM in Mar.'18 \cite{Temminghoff2019}.

Both of the detector arrays are surrounded by a multi-layer passive shield that consists of radiopure materials such as electroformed copper and ultra-low activity lead.
Furthermore, the inner shield is enclosed in an air-tight sealed box that is constantly flushed with evaporated dry nitrogen to suppress radon-induced radioactive backgrounds.

The raw data collected by the data acquisition (DAQ) consist of the digitized detector signals using flash analog-to-digital converters (FADCs) with a sampling frequency of $100\,$MHz and a sampling period of 10\,$\mu$s \cite{Cobra_demonstrator_2016}.
This enables a complex offline reconstruction and the identification of background-like events by means of pulse-shape discrimination \cite{Fritts2014, Zatschler2016}.
One of the key instruments regarding background suppression for COBRA is the reconstruction of the so-called interaction depth $z$, which is defined as the normalized distance between the CPG anode ($z=0$) and the planar cathode ($z=1$) \cite{Fritts2013}.

Besides COBRA's primary goal of searching for ultra-rare $\beta\beta$-decay transitions, one of the isotopes of interest present in CZT is \cd113 with a natural abundance of \mbox{12.227\% \cite{Iupac2016}} and an experimentally determined half-life of about $8\times 10^{15}$\,yr \cite{Cobra_Cd113_2009, Belli2007}.
Its fourfold forbidden non-unique $\beta$-decay has been investigated with different objectives in the past -- using the \mbox{COBRA} demonstrator \cite{Cobra_demonstrator_2016} and its predecessors, as well as CdWO$_4$ scintillator crystals (see e.g.~\cite{Cobra_Cd113_2020} for a comprehensive overview).

\section{Data-taking and event selection}
A detailed overview of the low-threshold \cd113 data-taking with the COBRA demonstrator as well as the detector calibration and characterization procedure can be found in \cite{Cobra_Cd113_2020}.
In order to ensure a stable operation during the dedicated \cd113 run, which lasted from Jul.'17 to Feb.'18, only a subset of the 64 CPG-CZT detectors of the demonstrator array has been taking data during this period.
Out of those detectors, 44 devices were selected for the data evaluation presented in this article.

In total, an isotopic exposure of 2.89\,kg\,d has been collected with an average energy threshold of 92\,keV, whereas the individual detector thresholds range from 52\,keV to 132\,keV.
The thresholds have been carefully optimized based on a dedicated noise study for each detector.
They ensure that there is no distortion of the spectrum-shape at low energies.
In addition, the detectors were kept at a constant temperature of about 9$^\circ$C.
This measure greatly reduced the thermal noise component and allowed for a stable operation at reasonably low thresholds.

Furthermore, a low-energy background model based on GEANT4 Monte-Carlo (MC) simulations and inputs from radiopurity assays of the detector construction materials has been developed as reported in \cite{Cobra_Cd113_2020}.
By comparing the model prediction and the combined COBRA data in the \cd113 energy range, a beneficial signal-to-background ratio of $S/B \sim 47$ could be reported, which indicates the overwhelming dominance of the \cd113 decay rate at low energies for the COBRA demonstrator.

An event based data selection is applied to the data, making use of several quality criteria in order to identify and reject unphysical pulse traces as well as events near the electrodes, which are known to be distorted by reconstruction artifacts \cite{Fritts2013}.
The validity and efficiency of those selections has been checked thoroughly with the help of both calibration and additional low-background physics data \cite{Zatschler2020}.
Moreover, it could be shown that the signal acceptance is sufficiently constant over the \cd113 energy range of interest, making it possible to compare the normalized experimental data with the predicted template spectra, and hence to perform the spectrum-shape analysis.

\section{Spectrum-shape analysis}
The spectrum-shape analysis follows the procedure outlined in \cite{Cobra_Cd113_2020, Zatschler2020} and will be summarized briefly in the following sections.
Its key feature is the comparison of theoretically calculated template spectra with the ones extracted from the COBRA data.
In the present case, a two-staged approach is being applied to first determine the best fit s-NME for each of the nuclear models using the combined COBRA data (see section~\ref{sec:sNME_fit}).
Secondly, the spectrum-shape analysis is performed for each individual detector spectra using the newly calculated templates based on the determined best fit s-NME for each nuclear model (see section~\ref{sec:gA_values}).
In both analysis steps, the respective template spectra as well as the experimental data must undergo several preparation steps.
Only then it is possible to perform a comparison of the normalized spectrum shapes.
These common steps are explained in the first two of the following sections.

\subsection{Template and data preparation}
In order to compare the experimental $\beta$-spectra recorded by the COBRA demonstrator array and the predicted template spectra based on the nuclear model calculations for fixed values of \gA, several preparation steps are needed.
The original template calculations have been carried out for $g_\text{A} \in [0.6,1.4]$ in steps of 0.01 with an energy binning of about 1\,keV.
A spline interpolation technique is used to produce template spectra for arbitrary \gA values in the given range with high accuracy.
Secondly, the detectors' finite energy resolution and the electron detection efficiency have to be accounted for in order to make the measured and theoretical spectra comparable.
Both effects are taken into account by folding the interpolated template spectra with the detector-specific energy resolution and the energy-dependent detector response function $\varepsilon_\text{det}(E)$.
The latter one has been determined via MC simulations for mono-energetic electrons that were homogeneously distributed over the full 1\,cm$^3$ crystal volume of a single CPG-CZT detector.

Finally, the experimental data as well as the prepared set of interpolated template spectra are normalized by the integral over the accessible energy range.
This range is limited by each detectors' optimized threshold and the \cd113 $\beta$-decay $Q$-value of $323.83 \pm 0.27$\,keV \cite{AME2016}.
In this way, it is possible to compare the spectrum-shapes for different values of \gA, independently of the total decay rate and the individual detector thresholds.

\subsection{Spectrum-shape comparison}
\label{sec:shape_comparison}
The degree of agreement between the experimental data and the predicted template spectra in dependence on \gA is evaluated with a standard $\chi^2$ test.
By taking into account the normalized decay rates $m_i$ for the energy bins $i$ to $N$,  including Poisson uncertainties $\sigma_i$ and the corresponding predictions $t_i$ for the same energy bin, the quantity $\chi^2$ is derived as
\begin{eqnarray}
\chi^2 = \sum \limits_{i=1}^N \left( \frac{m_i - t_i}{\sigma_i} \right)^2 .
\end{eqnarray}

In both the measured spectra and the predicted templates, the bin width is set to 4\,keV.
This width is a compromise between a preferably large number of bins $N$ and the resulting bin uncertainties arising from the number of bin entries.
For the average threshold of 92\,keV, the resulting number of bins is $N\sim 56$.

Initially, this method is applied to the combined spectrum of all detectors, using average parameters for the template preparation and taking into account the MC background model.
From the minima of the respective $\chi^2_\text{red}(g_{\rm A})$ curves a best match \gA value can be extracted.
The results are used to derive a best fit value for the s-NME that enters the template calculation for each of the nuclear models in the single detector analysis.

\begin{figure}[ht!]
\centering
\includegraphics[width=0.52\textwidth]{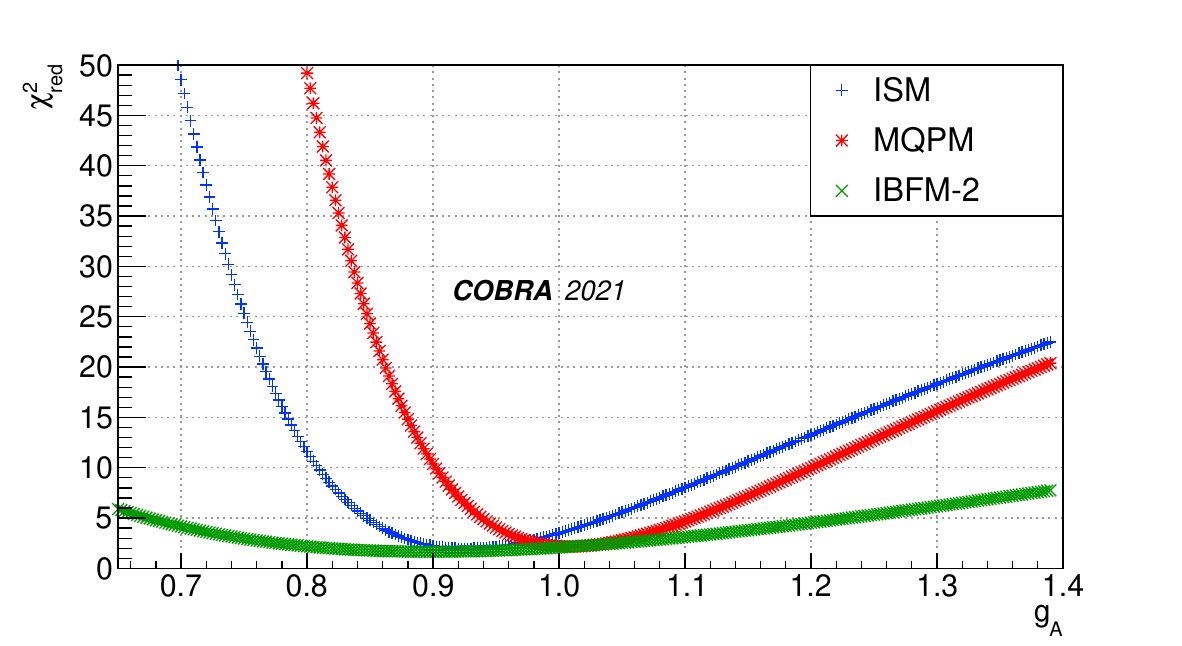}
\caption{Example $\chi^2_\text{red}(g_\text{A})$ curves of the spectrum-shape comparison between the spectrum of a single \mbox{COBRA} detector and the interpolated templates based on the ISM (blue), MQPM (red) and IBFM-2 (green) calculations. The shape of the $\chi^2_\text{red}(g_\text{A})$ curves and their features are representative for all single detector spectra of the selected ensemble. The analysis of the combined spectrum of all detectors has a similar outcome.}
\label{fig:chi2_curve}
\end{figure}

By applying the $\chi^2$ test to each of the 44 independent detector spectra and the corresponding sets of \gA templates, 44 best match \gA values are obtained.
A representative example of the $\chi^2_\text{red}(g_{\rm A})$ curves of a single detector analysis is depicted in Fig.~\ref{fig:chi2_curve}.
Compared to the original results reported in \cite{Cobra_Cd113_2020}, it turns out that the overall \gA dependence for the revised SSM is slightly reduced.
The uncertainty on each best match \gA is derived from the minimum $\chi^2 + 1$ as $1\sigma$ deviation, using a second order polynomial fit around the minimum.

The combined spectrum results serve as an internal cross-check for the individual detector results and allow to access some systematic uncertainties of the primary single detector analysis.
An example for those uncertainties is the neglection of the background modeling on the individual detector basis.

\subsection{Determination of s-NME}
\label{sec:sNME_fit}
\begin{figure*}[ht!]
\centering
\includegraphics[width=0.33\textwidth]{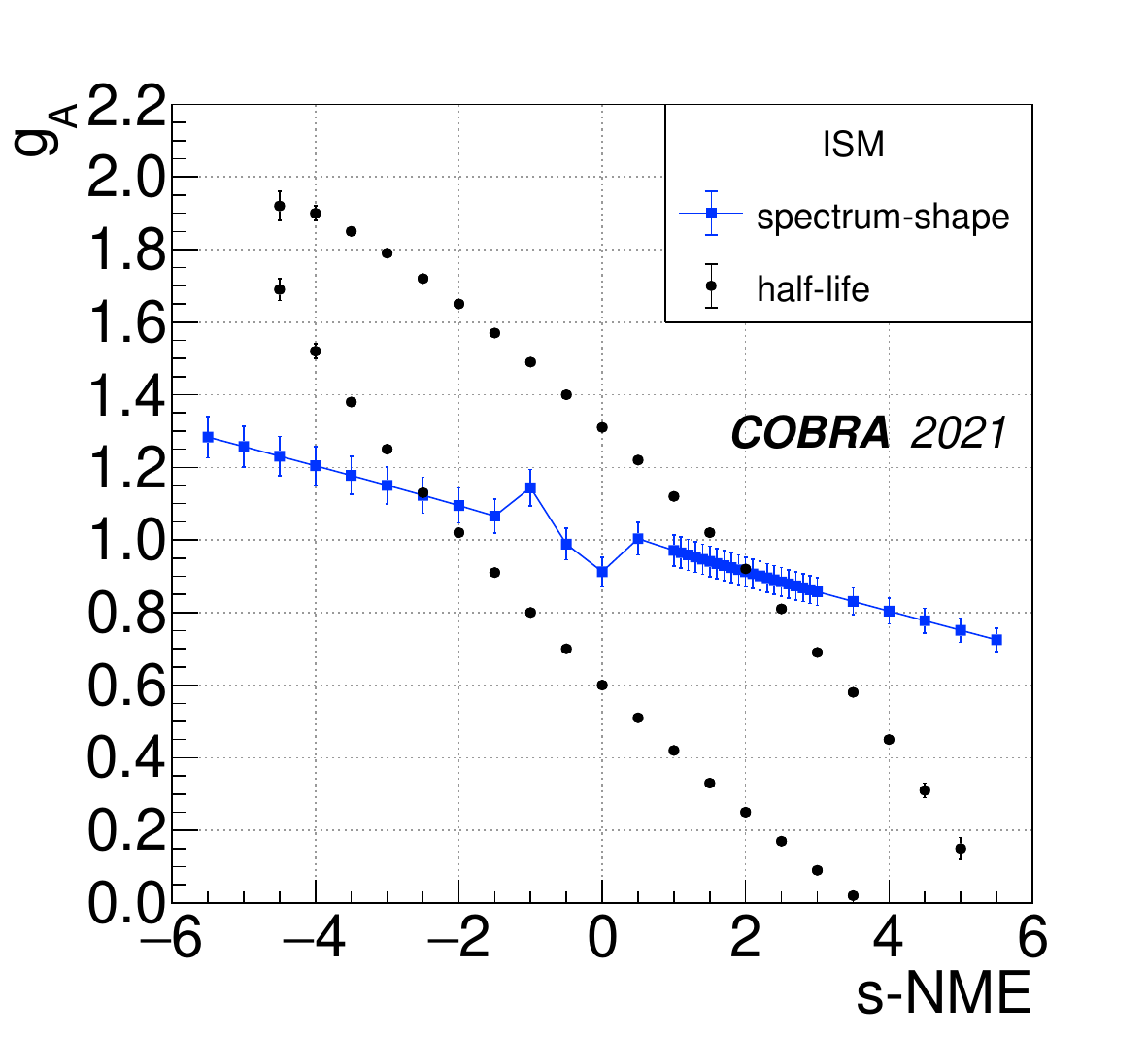}
\includegraphics[width=0.33\textwidth]{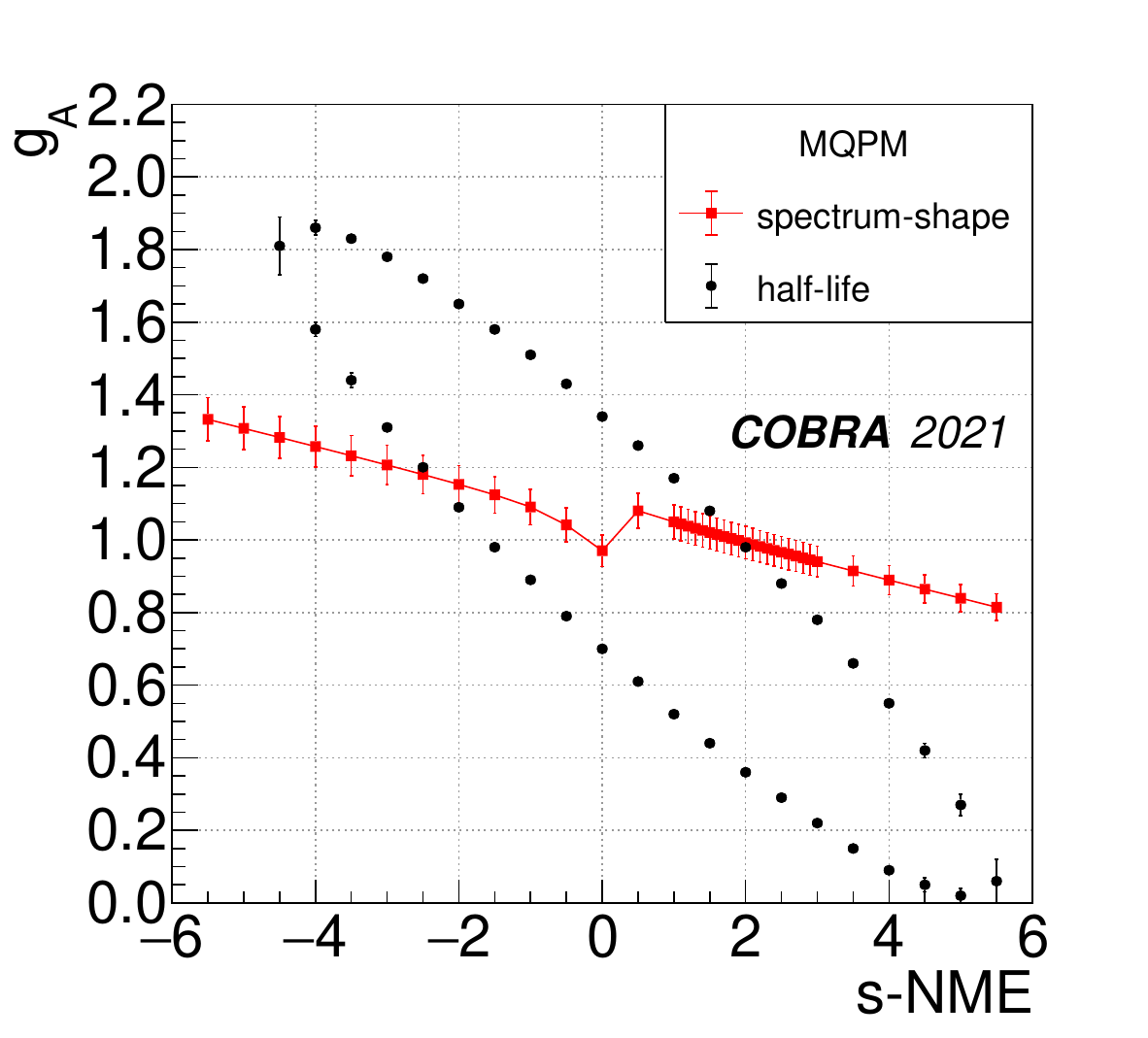}
\includegraphics[width=0.33\textwidth]{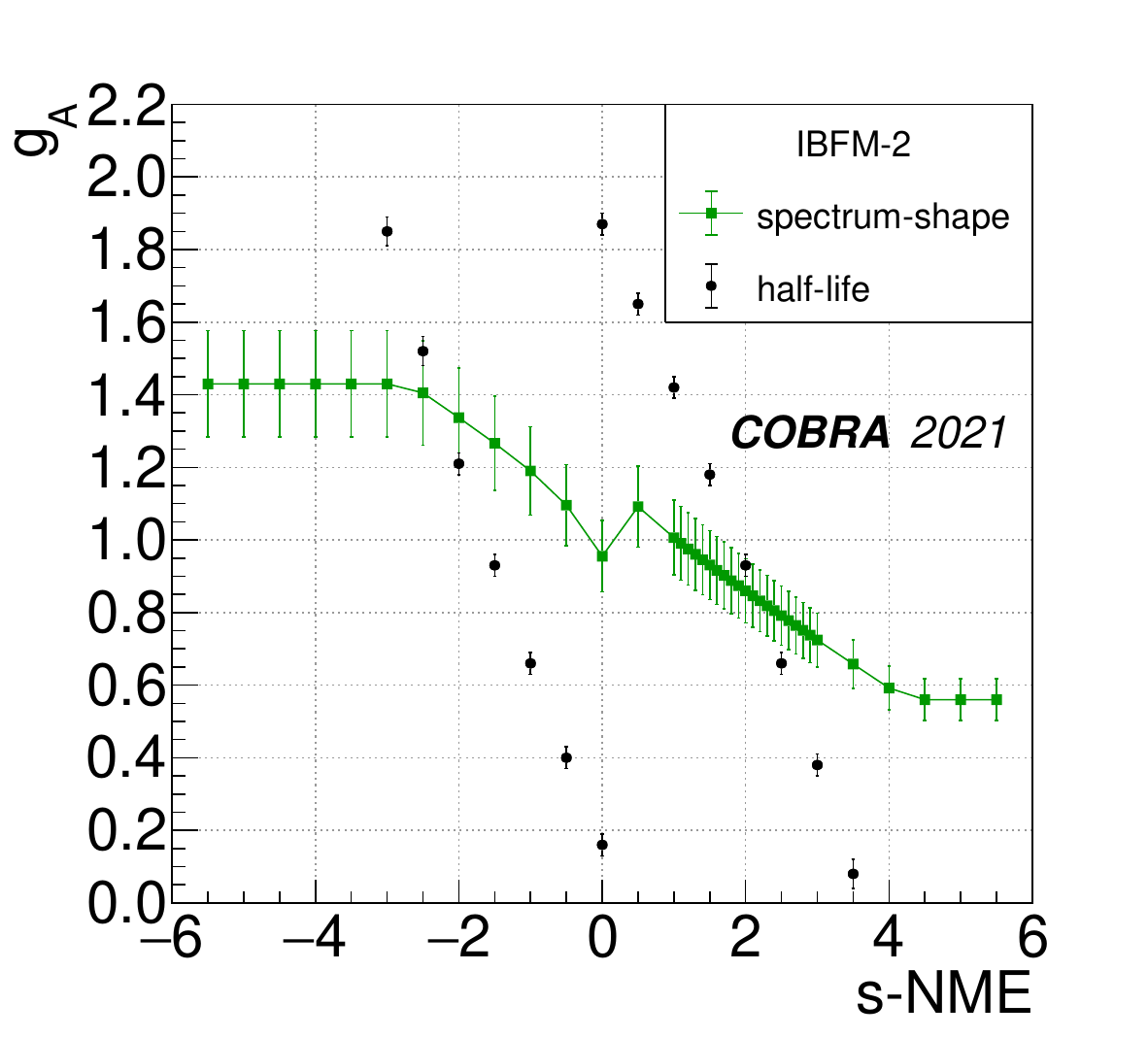}
\caption{Determination of the optimal s-NME values using the $g_{\rm A} (\text{s-NME})$ correlation resulting from the spectrum-shape analysis (color) and the half-life calculations (black). \textit{Left:} ISM. \textit{Middle:} MQPM. \textit{Right:} IBFM-2. For the half-life method the uncertainty on the single points arises from the range of values that is compatible with the quoted \cd113 reference half-life on the $2\sigma$ level. Only the data points along the curves for $\text{s-NME} > 0$ are assumed to be of physical relevance.}
\label{fig:gA_sNME}
\end{figure*}

As outlined in section~\ref{sec:CVC_improvement}, the s-NME remains as a free parameter in the CVC-inspired improvement of the SSM framework.
However, its value can be related to the half-life of the \cd113 $\beta$-decay in dependence on \gA.
By iterating \mbox{$\text{s-NME} \in [-5.5,+5.5]$} in steps of half a unit, the range of \gA values that reproduces the experimentally known \cd113 half-life within $2\sigma$ of the experimental uncertainties for a given value of the s-NME can be derived.
As half-life reference the most precise literature value based on a direct measurement $T_{1/2} = (8.04 \pm 0.05)\times 10^{15}$\,yr \cite{Belli2007} is used, whereas a similar value of $T_{1/2} = (8.00 \pm 0.26)\times 10^{15}$\,yr \cite{Cobra_Cd113_2009} has been obtained by a previous COBRA study.
This first step in the determination of the best fit s-NME results in an elliptic $g_{\rm A} (\text{s-NME})$ correlation as displayed in Fig.~\ref{fig:gA_sNME} for each of the three nuclear models.

Secondly, the spectrum-shape comparison is performed for each point of the two-dimensional \mbox{$\text{s-NME} \times g_{\rm A}$} parameter space using the combined data of the single COBRA detectors.
This step greatly reduces the number of $\chi^2(g_{\rm A})$ curve computations and is well-justified by the conformity of the average results presented in \cite{Cobra_Cd113_2020}.
Consequently, each \mbox{s-NME} input value can be associated with a best match \gA as shown in Fig.~\ref{fig:gA_sNME}.

Based on the outcome of the CVC-inspired calculations, resulting in small positive s-NME values for each of the nuclear models, the parameter space is restricted to $\text{s-NME} > 0$ in the following.
The intersection of the two $g_{\rm A}(\text{s-NME})$ curves in this range is taken as each model's best fit value of the s-NME.
Moreover, the uncertainty on the extracted effective \gA values using the combined \cd113 data is used to derive an uncertainty on these best fit values.
As a summary, Tab.~\ref{tab:NME_values} provides a numerical comparison of the l-NME appearing in Eqn.~(\ref{eqn:sNME}) and the \mbox{s-NME} as obtained in the original nuclear structure calculations, as well as for the CVC-inspired calculation and the derived best fit in combination with the spectrum-shape analysis.

\begin{table}[ht!]
\centering
\caption{Numerical values of the NMEs driving the spectrum-shape calculations for \cd113. The l-NME shown in the first column can be evaluated directly. For the calculation of the s-NME three different scenarios are presented. In each of the cases, the calculations were tuned to reproduce the \cd113 reference half-life. While in the first two s-NME approaches \gA needs to be fixed to the values shown in parentheses, it remains as a free parameter in the third one. This makes it possible to perform a simultaneous fit of the s-NME and \gA.}
\label{tab:NME_values}
\begin{small}
\begin{tabular}{@{\extracolsep{-2pt}}cccccc}
\toprule
 & l-NME [fm$^4$] & & \multicolumn{3}{c}{s-NME (\gA) [fm$^3$]} \\
\cline{2-2} \cline{4-6}
\addlinespace
model  & $^V\mathcal{M}^{(0)}_{440}$ & & unadj. & CVC & best fit \\
\midrule
ISM    & 719.3 & & 0 (1.29)    & 0.53 (1.20) & $1.97 \pm 0.21$ \\
MQPM   & 827.1 & & 0.37 (1.32) & 0.61 (1.18) & $1.85 \pm 0.19$ \\
IBFM-2 & 316.8 & & 0 (1.89)    & 0.23 (1.80) & $2.10 \pm 0.17$ \\
\bottomrule
\end{tabular}
\end{small}
\end{table}

It should be noted that for the original and the CVC-inspired calculations it is necessary to fix \gA to a range where the half-life is reproduced within a feasible s-NME range.
This restriction leads to $g_{\rm A} \sim 1.2 - 1.8$.
In contrast to that, the novel approach devised in the present work allows to fit the s-NME values that reproduce the known half-life in combination with an effective \gA resulting from the spectrum-shape analysis for each of the nuclear models.
The best fit values of the s-NME $^{V}\mathcal{M}^{(0)}_{431}$ are reported in Tab.~\ref{tab:NME_values}.

A priori, similar values of the s-NME for the three nuclear models are not expected.
This stems from the fact that the three models are quite different in their basic assumptions of simplifying the nuclear many-body problem.
The MQPM has a large valence (single-particle) space but restricted many-body configurations.
The ISM has a very restricted valence space but a complete set of many-body configurations.
The IBFM-2 has a similarly restricted valence space as the ISM and a similarly restricted many-body configuration space as the MQPM.
In this respect the IBFM-2 is the most schematic model with a very strong phenomenological renormalization of its many parameters.
These aspects are reflected in the differences between the CVC-determined and fitted values of the s-NME for each model: the smallest difference is found for the MQPM and the largest for the IBFM-2, and only the MQPM is able to produce a non-zero value of the s-NME just from the nuclear-structure calculation.
An improved nuclear model could combine the assets of the ISM and MQPM -- a large valence space and a complete configuration space.
Then it is expected that the difference between the mentioned two s-NME values would be minor.

Finally, each nuclear model's best fit s-NME enters the calculation of an optimized set of template spectra to address the quenching of \gA in the full single-detector spectrum-shape analysis following the steps outlined in section~\ref{sec:shape_comparison}.

\subsection{Effective \gA values}
\label{sec:gA_values}
\begin{figure}[ht!]
\centering
\includegraphics[width=0.52\textwidth]{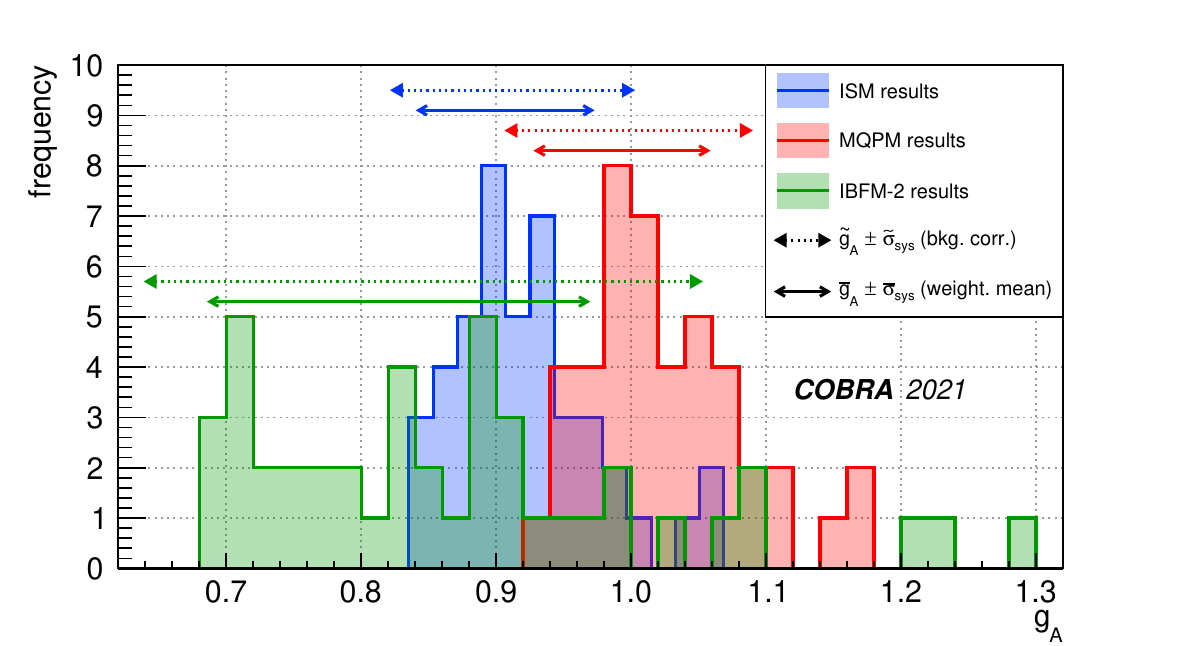}
\caption{Distribution of the 44 best match \gA values for the ISM (blue), MQPM (red) and IBFM-2 (green) calculations. Furthermore, the respective weighted means $\overline{g}_\text{A} \pm \overline{\sigma}_\text{sys}$ as well as the results of the spectrum-shape comparison for the combined experimental data $\tilde{g}_\text{A} \pm \tilde{\sigma}_\text{sys}$ including the background correction are highlighted. The spread of the \gA distributions agrees well with the determined systematic uncertainties on the mean values.}
\label{fig:gA_distribution}
\end{figure}

The distributions of the best match \gA values resulting from the $\chi^2$ test for the 44 independent single detector measurements are shown in Fig.~\ref{fig:gA_distribution}.
In contrast to the ISM and MQPM results, which are rather tightly distributed, the IBFM-2 distribution turns out to be rather flat.
This is due to the fact that the latter model is less sensitive to \gA as could already be seen in the corresponding $\chi^2_\text{red}(g_{\rm A})$ curve in Fig. \ref{fig:chi2_curve}.

By combining the single detector results into a weighted mean, an average $\overline{g}_{\rm A}$ for each of the three nuclear models can be derived.
This is done by using the $\chi^2+1$ deviation from the minimum of the $\chi^2(g_{\rm A})$ curve as the weight and a measure of the statistical uncertainty.
It turns out that the absolute statistical uncertainty from the sum of the weights is on the order of $\overline{\sigma}_\text{stat} \sim 1\cdot 10^{-3}$ for the ISM and MQPM and about four times higher in case of the IBFM-2.
In comparison to the derived systematic uncertainties $\overline{\sigma}_\text{sys}$, which will be summarized in section~\ref{sec:systematics}, it is possible to neglect those statistical uncertainties in the final results.
The extracted weighted means including the dominating systematic uncertainties are
\begin{eqnarray}
\overline{g}_{\rm A}(\text{ISM})    &=& 0.907 \pm 0.064, \label{eqn:gA_eff01}\\
\overline{g}_{\rm A}(\text{MQPM})   &=& 0.993 \pm 0.063,\\
\overline{g}_{\rm A}(\text{IBFM-2}) &=& 0.828 \pm 0.140. \label{eqn:gA_eff03}
\end{eqnarray}

These values agree also well with the results obtained for the combined spectrum analysis.
However, because the latter procedure makes use of averaged parameters in the template preparation, its systematic uncertainties are about 30\% higher than for the single detector analysis.
This is why the single detector analysis, which allows for a much more nuanced treatment of the individual detector effects, is chosen as the primary analysis strategy, while the combined spectrum analysis serves as a cross-check and is used to determine the optimal s-NME values.
A compilation of the single detector spectra as well as their corresponding best match \gA templates can be found in Fig.~\ref{fig:appendix_single_detector_results01}\,--\,\ref{fig:appendix_single_detector_results04} in the appendix.

\subsection{Evaluation of systematic uncertainties}
\label{sec:systematics}
The list of considered systematic uncertainties is based on our previous study presented in \cite{Cobra_Cd113_2020}.
It consists of parameters relevant to the spectrum-shape analysis as well as the fitting of the s-NME values that enter the template calculations.
Each parameter has been evaluated separately by modifying only one parameter at a time within conservative limits while all others are fixed to their default values throughout the analysis.
The modulus of the difference between the altered and the default \gA results are then taken as a measure for the systematic uncertainty.
None of the considered parameters turns out to be negligible as they all lead to deviations which are significantly larger than a corresponding $3\sigma$ deviation expected from solely statistical fluctuations.
The total systematic uncertainty for each model is obtained as the square root of the sum of squared uncertainties.
A summary of the systematics and their corresponding impact on the average \gA results is given in Tab.~\ref{tab:systematics}.

\begin{table}[ht!]
\caption{List of systematic uncertainties considered in the \cd113 spectrum-shape analysis under the revised SSM and CVC hypothesis.}
\label{tab:systematics}
\begin{small}
\begin{tabular}{ccccc}
\toprule
          &  & uncertainty [\%] &  \\
parameter & ISM & MQPM & IBFM-2\\ 
\midrule
template interpolation        & 0.001 & 0.001 & 0.001\\
detector response $\varepsilon_\text{det}(E)$ & 0.051 & 0.037 & 0.135\\
resolution FWHM$(E)$          & 0.201 & 0.156 & 0.072\\
energy calibration		      & 5.285 & 4.380 & 15.107\\
analysis threshold		      & 3.978 & 4.123 & 2.340\\
$z$-cut selection		      & 1.626 & 1.325 & 5.074\\
$\chi^2_\text{red}$ fit range & 0.005 & 0.008 & 0.009\\
background modeling			  & 0.042 & 0.039 & 0.435\\
s-NME determination			  & 1.833 & 1.616 & 5.009\\
\midrule
total 						  & 7.054 & 6.370 & 16.874\\
\bottomrule
\end{tabular}
\end{small}
\end{table}

The largest contribution arises from the uncertainty on the energy calibration, which has been determined to \mbox{$\Delta E = \pm 1.3$\,keV} for the \cd113 energy range.
Following this, the analysis threshold and the fiducial volume selection based on the reconstructed interaction depth $z$ lead to additional contributions of similar magnitude.
A comparable contribution also arises from the uncertainty on the best fit s-NME values.
In total, the systematic uncertainties add up to values on the few percent level and agree well with the observed spread of the \gA values extracted from the single detector analysis (see Fig.~\ref{fig:gA_distribution}).

Compared to the previous results reported in Ref.~\cite{Cobra_Cd113_2020}, the overall uncertainty on $\overline{g}_{\rm A}$ increased quite significantly by about five times.
This is a direct consequence of the weakened \gA dependence of the underlying templates that follow from the nuclear model calculations.
Despite this, the determined effective $\overline{g}_{\rm A}$ values in Eqn.~(\ref{eqn:gA_eff01}) -- (\ref{eqn:gA_eff03}) are perfectly consistent with the previously reported results.

\subsection{Model preference}
\begin{figure*}[ht!]
\centering
\includegraphics[width=0.33\textwidth]{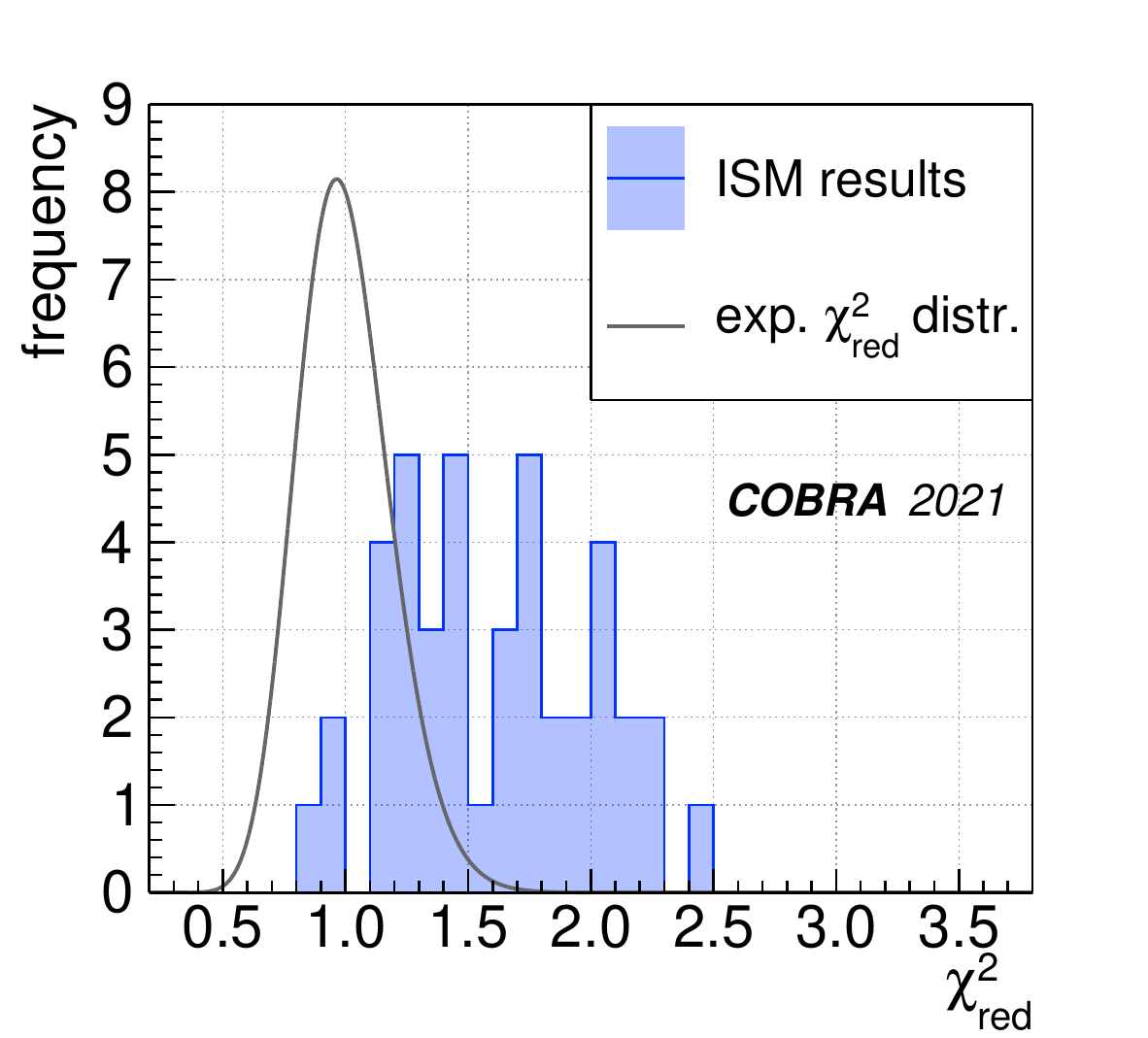}
\includegraphics[width=0.33\textwidth]{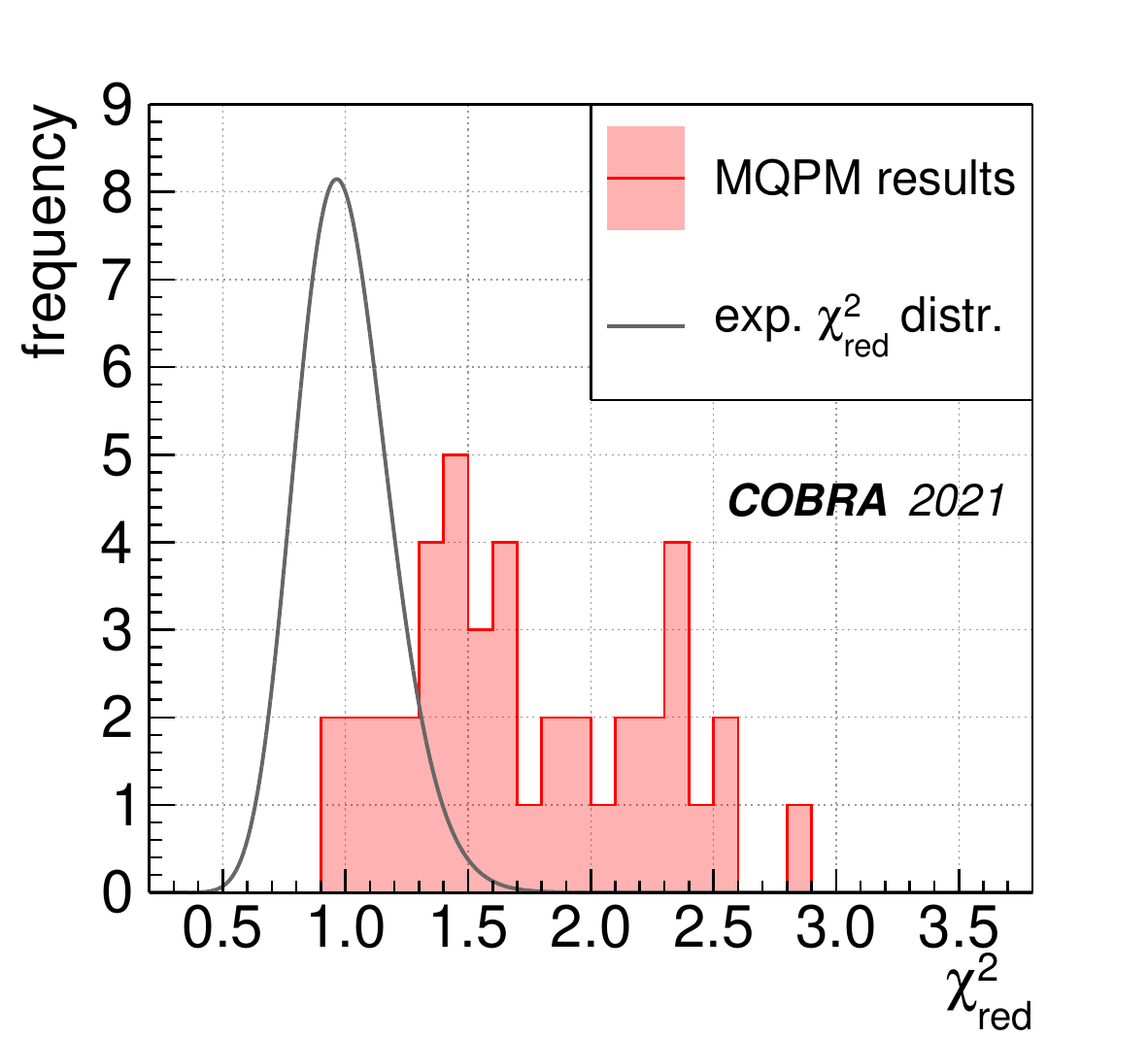}
\includegraphics[width=0.33\textwidth]{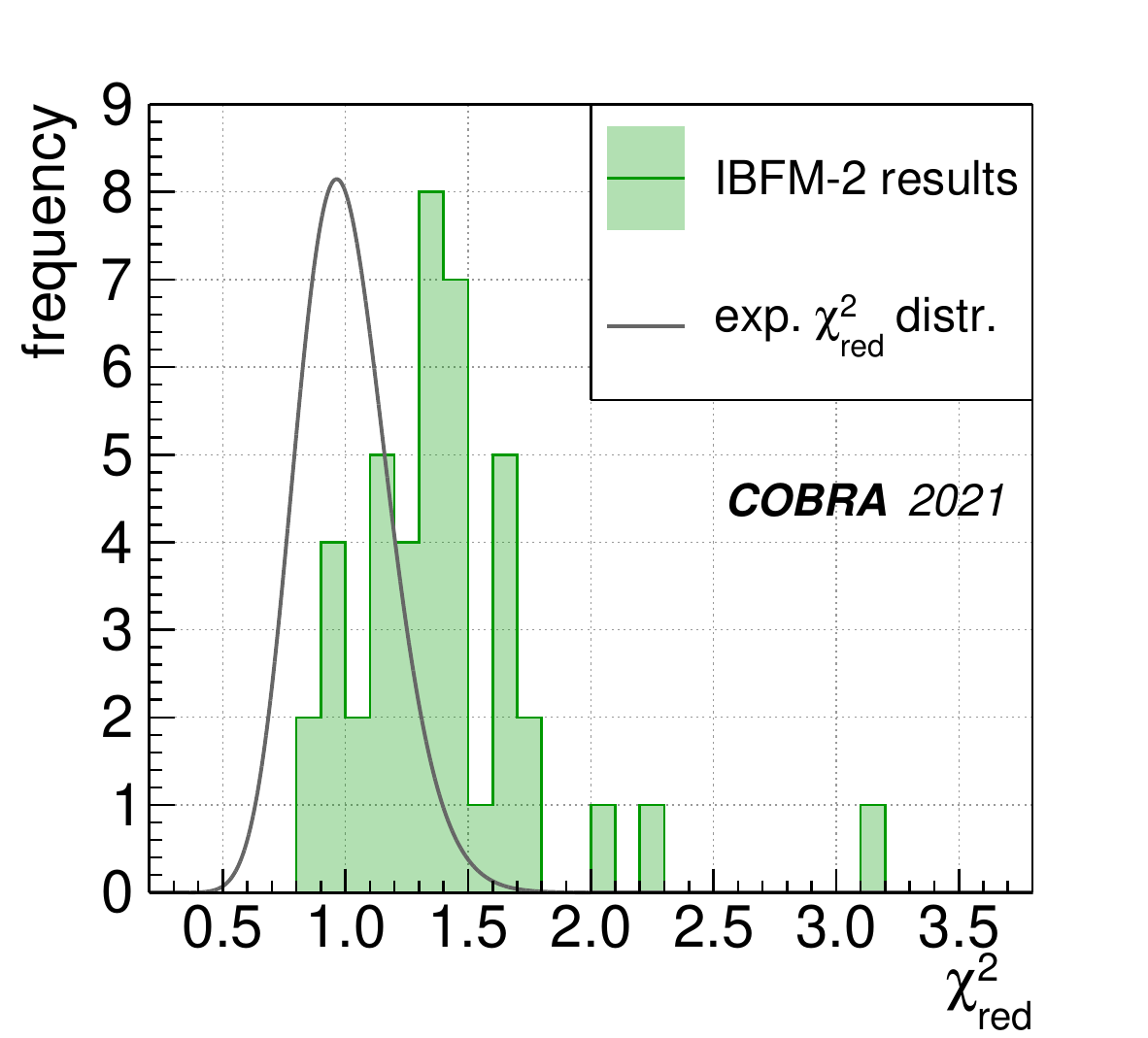}
\caption{Distribution of the minimum $\chi^2_\text{red}$ values of the best match \gA values for the single detector spectrum-shape comparison.
All three nuclear models result in similar distributions around $\overline{\chi}^2_\text{red} \approx 1.5$.
While this indicates a similar degree of agreement between the experimental data and the model predictions, most values do not comply with the expected $\chi^2_\text{red}$ distribution for the average number of degrees of freedom of about 56,  shown in grey.
This observation is an indication for potential systematic uncertainties, which were not accounted for, and that affect the $\chi^2$ test procedure as well as its statistical interpretation.}
\label{fig:chi2_distributions}
\end{figure*}

The distributions of the minimum $\chi^2_\text{red}$ values for the single detectors' best match \gA values can be taken as an indication for the agreement between the COBRA data and the provided nuclear model predictions.
They are shown in Fig.~\ref{fig:chi2_distributions}.

It is worth noting that the most prominent outliers can be identified as the same detector units, regardless of the chosen nuclear model.
The minimum $\chi^2_\text{red}$ distributions turn out to be very similar for the three considered nuclear models.
The distributions can be further characterized by determining their mean values and the respective uncertainties on the mean as summarized in Eqn.~(\ref{eqn:chi2_01}) -- (\ref{eqn:chi2_03}):
\begin{eqnarray}
\overline{\chi}^2_\text{red}(\text{ISM})    &=& 1.60 \pm 0.06, \label{eqn:chi2_01}\\
\overline{\chi}^2_\text{red}(\text{MQPM})   &=& 1.74 \pm 0.07, \\
\overline{\chi}^2_\text{red}(\text{IBFM-2}) &=& 1.35 \pm 0.05. \label{eqn:chi2_03}
\end{eqnarray}

These values indicate a similar agreement between the data and the different model predictions.
A similar outcome has been reported in our previous analysis \cite{Cobra_Cd113_2020}.
If one compares the $\chi^2_\text{red}$ distributions to the expected distribution given an average number of degrees of freedom of about 56, no clear model preference can be reported.

The deviations from the expected $\chi^2_\text{red}$ distribution are presumably caused by an underestimation of the systematic uncertainties related to the nuclear model calculations.
In the current implementation of the SSM it is not feasible to incorporate such model uncertainties due to the complexity of the involved NME calculations and the lack of systematic ways to assess the model uncertainties.
This issue might be resolved in the future with the advances made in the systematic error estimates of nuclear-model uncertainties.
In fact, steps towards this goal have been taken in the frameworks of the \textit{ab initio} nuclear many-body theory \cite{Koenig2014, Odell2016} and the energy density functionals \cite{Doba2014, Roca2015, Schunck2015a, Schunck2015b}.

\subsection{Shape factor}
\begin{figure*}[ht!]
\centering
\includegraphics[width=0.33\textwidth]{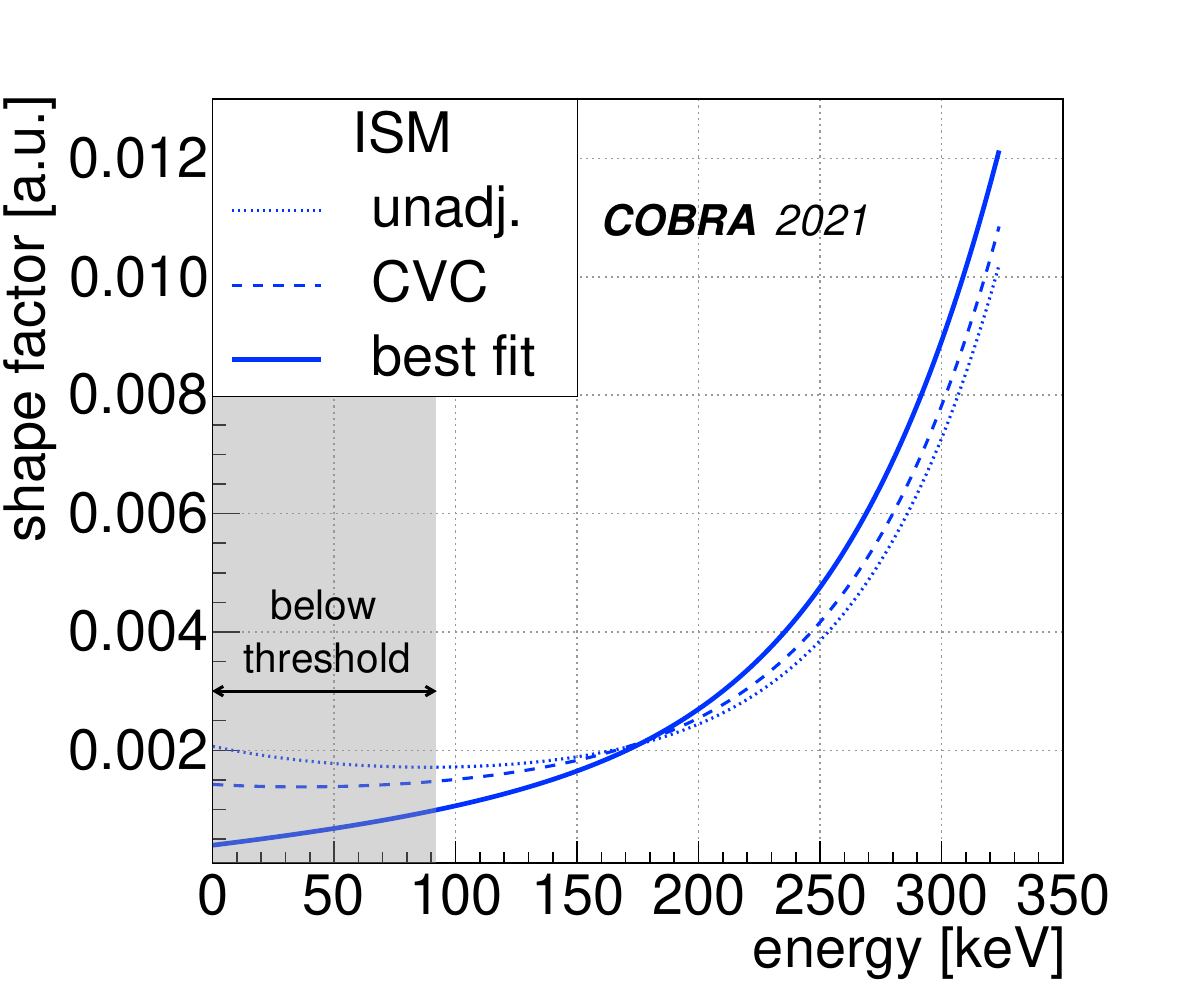}
\includegraphics[width=0.33\textwidth]{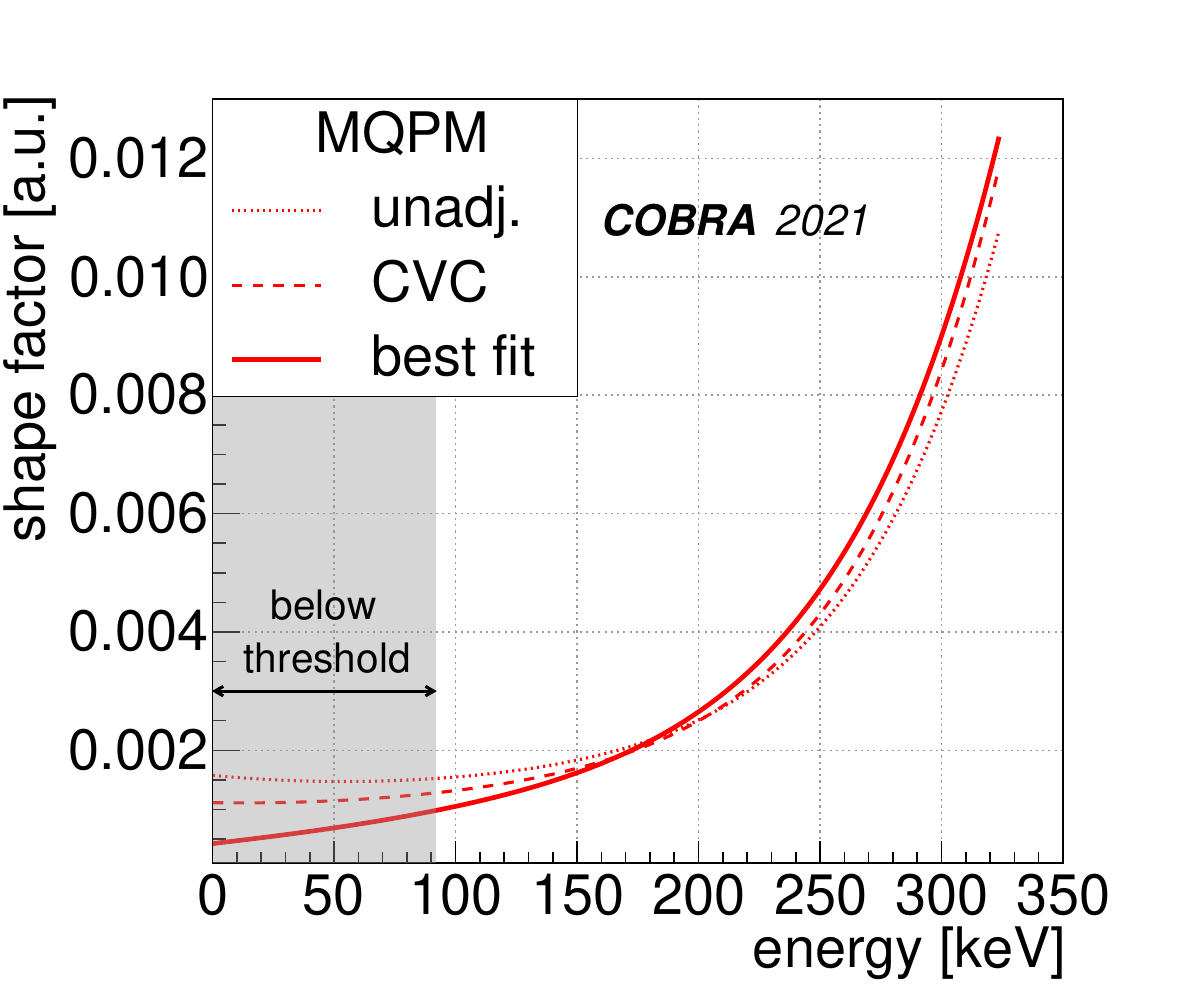}
\includegraphics[width=0.33\textwidth]{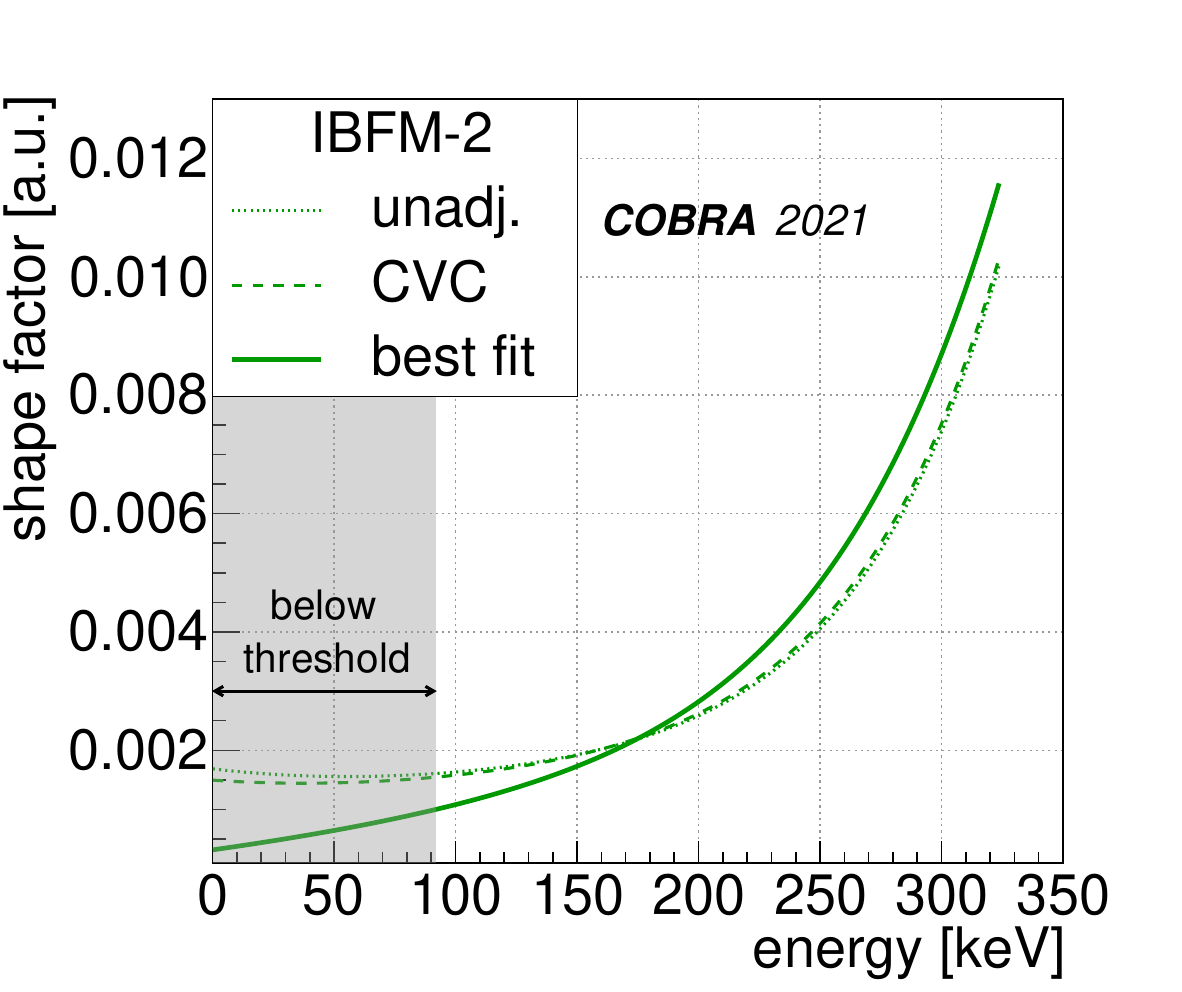}
\caption{Visualization of the \cd113 $\beta$-decay shape factor $C(w_e)$ for different scenarios. \textit{Left:} ISM. \textit{Middle:} MQPM. \textit{Right:} IBFM-2. Each of the shape factor curves is normalized to unity. The grey shaded area indicates the average experimental threshold of the COBRA detectors in the dedicated \cd113 low-threshold data-taking.}
\label{fig:Cd113_shape_factor}
\end{figure*}

The determined values of the effective $\overline{g}_\text{A}$ in combination with the best fit s-NME values reported in Tab.~\ref{tab:NME_values} can be used to illustrate the shape factor $C(w_e)$ introduced in Eqn.~(\ref{eqn:shape_factor}) that is associated with the \cd113 $\beta$-decay.
The results for each of the three nuclear models are depicted in Fig.~\ref{fig:Cd113_shape_factor}.
The overall energy dependency is similar for all of the displayed cases whereas the largest differences are observed for the low-energy range.
For the results corresponding to the best fit approach, the shape factors are found to agree very well between the three models.

The same applies to the normalized spectrum-shapes taking into account the determined systematic uncertainties on the effective $\overline{g}_\text{A}$ values reported in Eqn.~(\ref{eqn:gA_eff01})\,--\,(\ref{eqn:gA_eff03}).
An illustration can be found in Fig.~\ref{fig:Cd113_allowed_spectrum_range} of the appendix.
The visualization makes use of an interpolation based on splines of the original template spectra within the provided \gA range to predict templates for arbitrary \gA values.
The degree of agreement down to energies that are below the average analysis threshold of 92\,keV could not be reported in our previous \cd113 study \cite{Cobra_Cd113_2020}.
It is a direct consequence of the improved understanding of the implications of the CVC hypothesis and its impact on the NME calculation within the SSM.
Nonetheless, a lower experimental threshold would be beneficial in a future follow-up campaign to probe the spectrum-shape at low energies directly.

\subsection{Half-life reproducibility}
By adjusting the matrix elements that drive the spectrum-shape calculations in a way that reproduces the experimentally known half-life of the \cd113 $\beta$-decay, one of the severe shortcomings of the SSM that has been discussed in the past (see e.g.~\cite{Haaranen2017}) could be resolved.
In fact, based on the revised nuclear model calculations a correlation between the \cd113 half-life and the effective value of \gA can be derived which is unambiguous in the sense that for $g_\text{A} > 0.7$ there is only a single \gA pointing to a unique $T_{1/2}$ for each of the models and vice versa.

In order to access arbitrary half-life values in the given \gA range, the discrete data points $T_{1/2}(g_\text{A})$ from the calculations can be described analytically by a spline.
With the spline approximation it is then possible to convert the experimental $\overline{g}_\text{A} \pm \overline{\sigma}_\text{sys}$ results in Eqn.~(\ref{eqn:gA_eff01}) -- (\ref{eqn:gA_eff03}) into a respective half-life range.
This leads to the following values:
\begin{eqnarray}
T_{1/2}(\text{ISM})    &=& 8.6^{+3.8}_{-2.3} \times 10^{15}\,\text{yr}, \label{eqn:Cd113_gA_halflife01}\\
T_{1/2}(\text{MQPM})   &=& 6.9^{+2.9}_{-1.8} \times 10^{15}\,\text{yr}, \\
T_{1/2}(\text{IBFM-2}) &=& 10.3^{+4.2}_{-2.7} \times 10^{15}\,\text{yr}. \label{eqn:Cd113_gA_halflife03}
\end{eqnarray}

\begin{figure*}[ht!]
\centering
\includegraphics[width=0.33\textwidth]{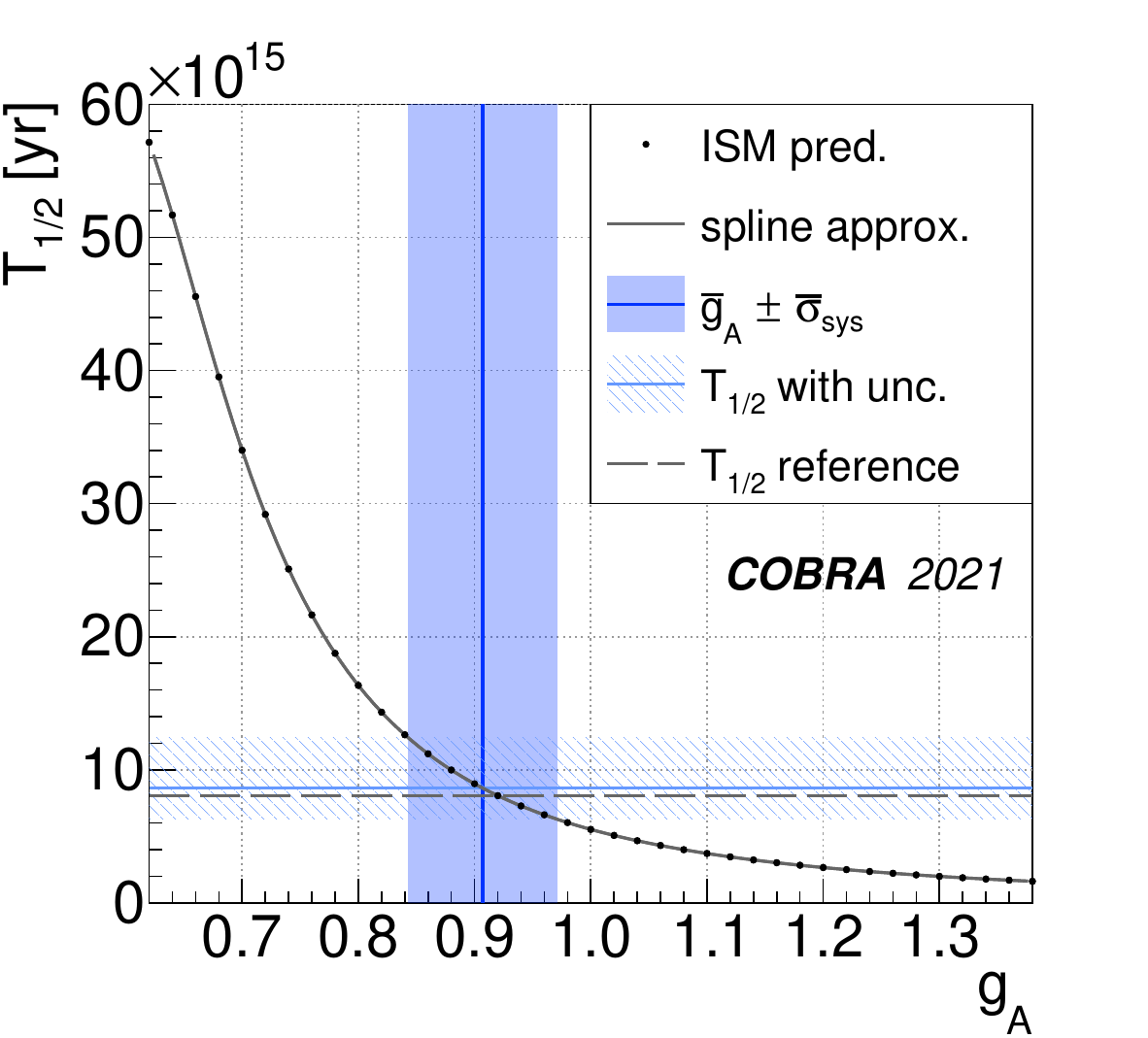}
\includegraphics[width=0.33\textwidth]{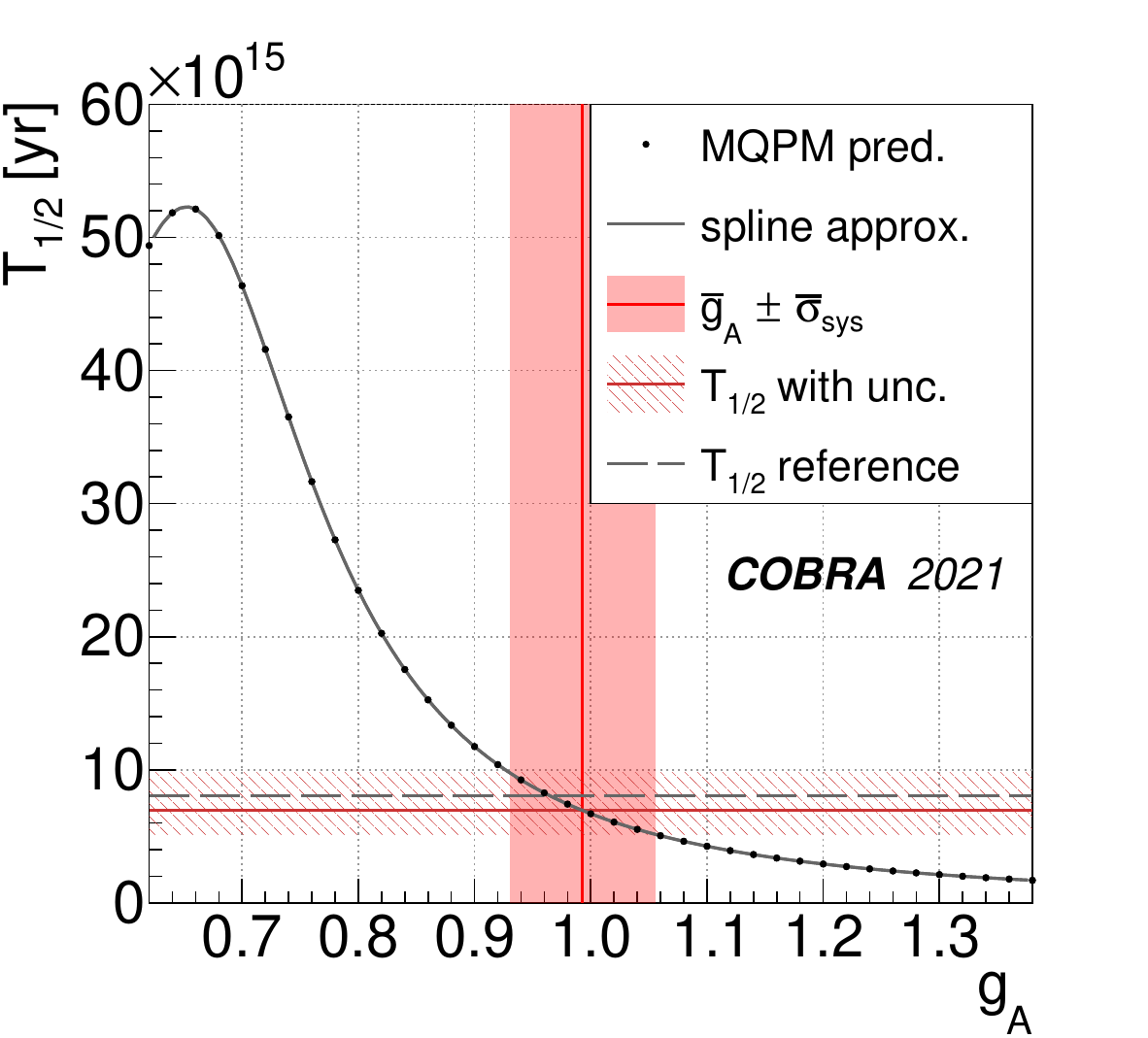}
\includegraphics[width=0.33\textwidth]{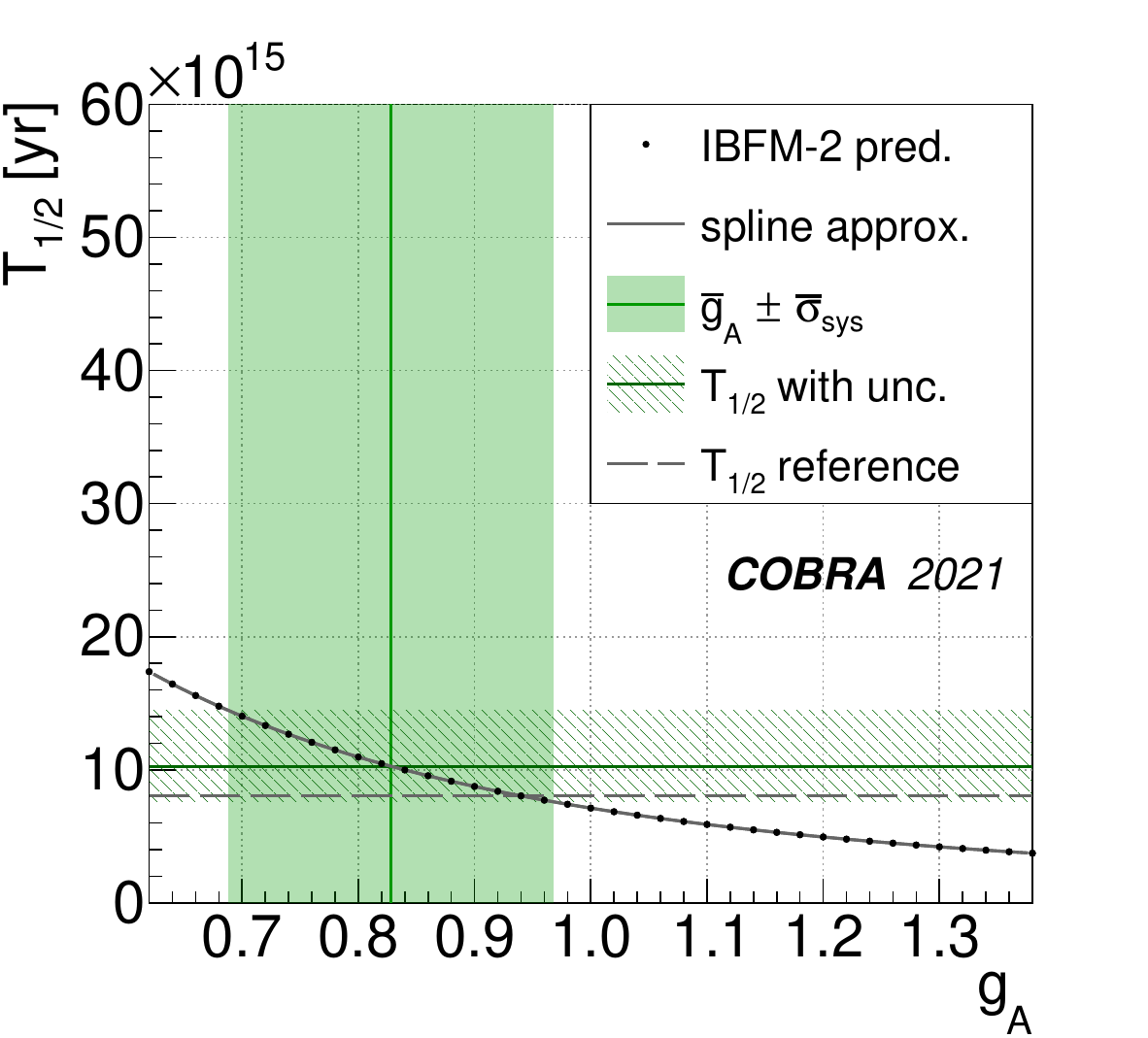}
\caption{Conversion of the experimental results $\overline{g}_\text{A} \pm \overline{\sigma}_\text{sys}$ into a corresponding half-life range following the revised SSM prediction under the CVC hypothesis. \textit{Left:} ISM. \textit{Middle:} MQPM. \textit{Right:} IBFM-2. All three nuclear model results are consistent with the \cd113 reference half-life of about $8.0 \times 10^{15}$\,yr.}
\label{fig:Cd113_gA_halflife}
\end{figure*}

An illustration of the predicted half-life in dependence on the effective \gA for all three nuclear models is shown in Fig.~\ref{fig:Cd113_gA_halflife}.
Only for the MQPM there is a small range of $0.6 < g_\text{A} < 0.7$ for which the same predicted half-life gets reproduced by multiple \gA values.
For higher \gA values close to the free one, the correlation function $T_{1/2}(g_\text{A})$ gets rather flat for all models.
The IBFM-2 prediction is again found to be the least sensitive to \gA.

The results summarized in Eqn.~(\ref{eqn:Cd113_gA_halflife01}) -- (\ref{eqn:Cd113_gA_halflife03}) illustrate how the uncertainties on \gA affect the \cd113 half-life determination by using solely the spectrum-shape.
It is remarkable that all of them cover the quoted \cd113 reference half-lives.
Due to the shape of the predicted $T_{1/2}(g_\text{A})$ correlation, the resulting uncertainties on the extracted half-lives turn out to be asymmetric around the central value.
In relative units, the uncertainties are on the order of $\sim 30\%$ for all models.
The half-life uncertainties are directly dependent on the precision of the extracted effective \gA values which in turn are dominated by several systematic effects as discussed in section~\ref{sec:systematics}.

Alternatively, it would be possible to determine the \cd113 half-life by performing an absolute rate measurement using the same data set.
However, this requires a more detailed understanding of the involved efficiencies and a more nuanced treatment of the individual detectors.
This level of understanding is not required for the present spectrum-shape study as all involved spectra have been normalized accordingly.
Moreover, if one wanted to determine the half-life by an absolute rate measurement, the total rate could only be determined by making an assumption about the shape of the $\beta$-spectrum below the accessible energy range.
For this, the allowed spectrum ranges as shown in Fig.~\ref{fig:Cd113_allowed_spectrum_range} in the appendix could provide some guidance.

\section{Summary and conclusion}
The present article addresses an analysis of the \cd113 $\beta$-decay based on a low-threshold data set collected by the COBRA demonstrator at the LNGS and several nuclear model calculations using the revised SSM under the improved CVC hypothesis.
The evaluation of the normalized experimental \cd113 spectra of 44 individual CPG-CZT detectors in the context of the applied nuclear models results in significantly quenched, effective values of the weak axial-vector coupling \gA.
The determined systematic uncertainties add up to about 6-7\% for the ISM and MQPM and to about 17\% for the IBFM-2 calculations.

One of the long-standing shortcomings of the SSM -- being unable to predict the \cd113 half-life and spectrum-shape consistently -- could be resolved by performing a data-driven fit of the small relativistic NME in combination with an effective value of \gA.
This achievement is an important step regarding the description of strongly forbidden $\beta$-decays such as that of \cd113.

As a final remark, it is to be remembered that the SSM is based on a subtle cancellation between the three terms in Eqn.~(\ref{eqn:shape_factor}). In the majority of $\beta$-decay transitions such cancellation does not happen and the SSM does not work. Nevertheless, it is conceivable that the SSM-determined \gA is appropriate for all low-energy weak processes of the same degree of forbiddenness. It is a matter of further studies to find out if the SSM determined \gA value is relevant for other low-energy weak processes such as the $2\nu\beta\beta$-decay, which is a combination of allowed Gamow-Teller transitions. Things standing as stated, it is of paramount importance to explore other $\beta$ transitions where the SSM is applicable.
For a recent compilation regarding the experimental confirmation of \gA quenching in weak interactions and an overview of potentially interesting nuclear decays see the review article \cite{Suhonen_report_2019}.
Among the list of nuclear decays presented there are non-unique as well as unique transitions with an angular momentum exchange up to $\Delta J = 4$.
The effect of an effectively quenched \gA on transitions with such a high angular momentum exchange is expected to be the most prominent, hence, could be used in follow-up campaigns to gather more experimental data on the subject of \gA quenching.

\section*{Acknowledgments}
The COBRA collaboration greatly appreciates the continuous support by the LNGS and the access to its infrastructure and resources.
Moreover, the collaboration would like to thank the German Research Foundation (Deutsche Forschungsgemeinschaft, DFG) for its past funding and the support of the COBRA XDEM upgrade under the project numbers ZU123/3 and GO1133/1.
Additionally, the support by the Academy of Finland under the project number 318043 and from the Jenny and Antti Wihuri Foundation is acknowledged.

\bibliographystyle{unsrt}
\bibliography{Cd113_gA_study_2021_PLB}

%\begin{thebibliography}{10}

%\end{thebibliography}

\appendix
\onecolumn
\section{Visualization of spectrum-shape dependence}

\begin{figure*}[ht!]
\centering
\includegraphics[width=0.7\textwidth]{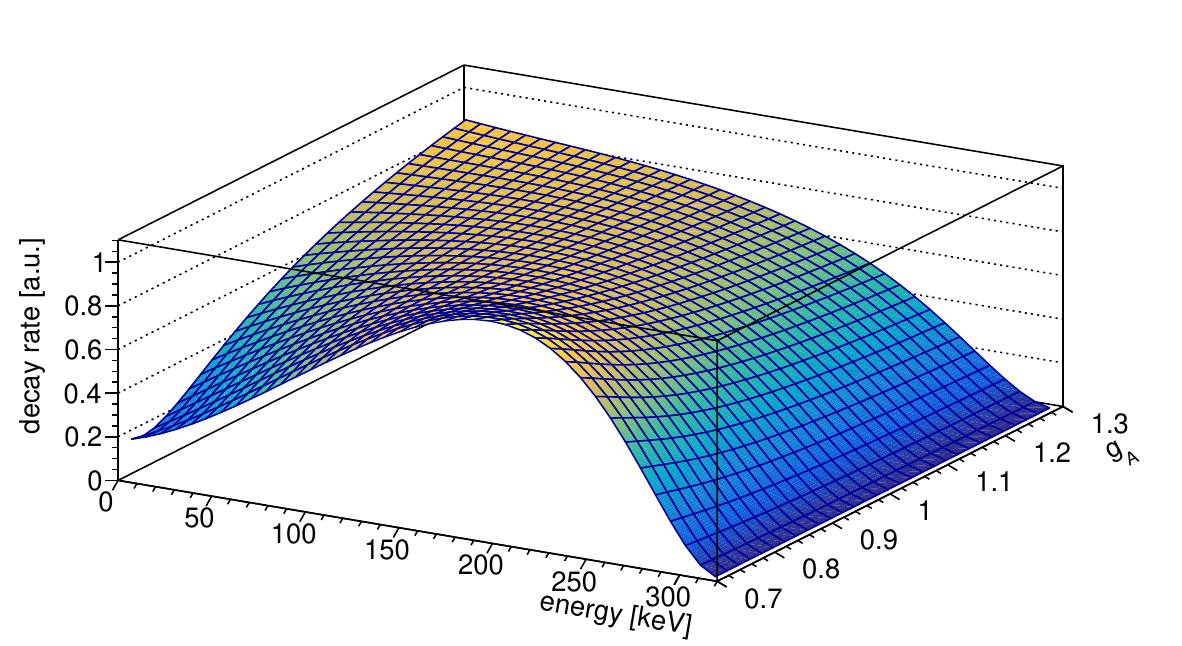}\\
\includegraphics[width=0.7\textwidth]{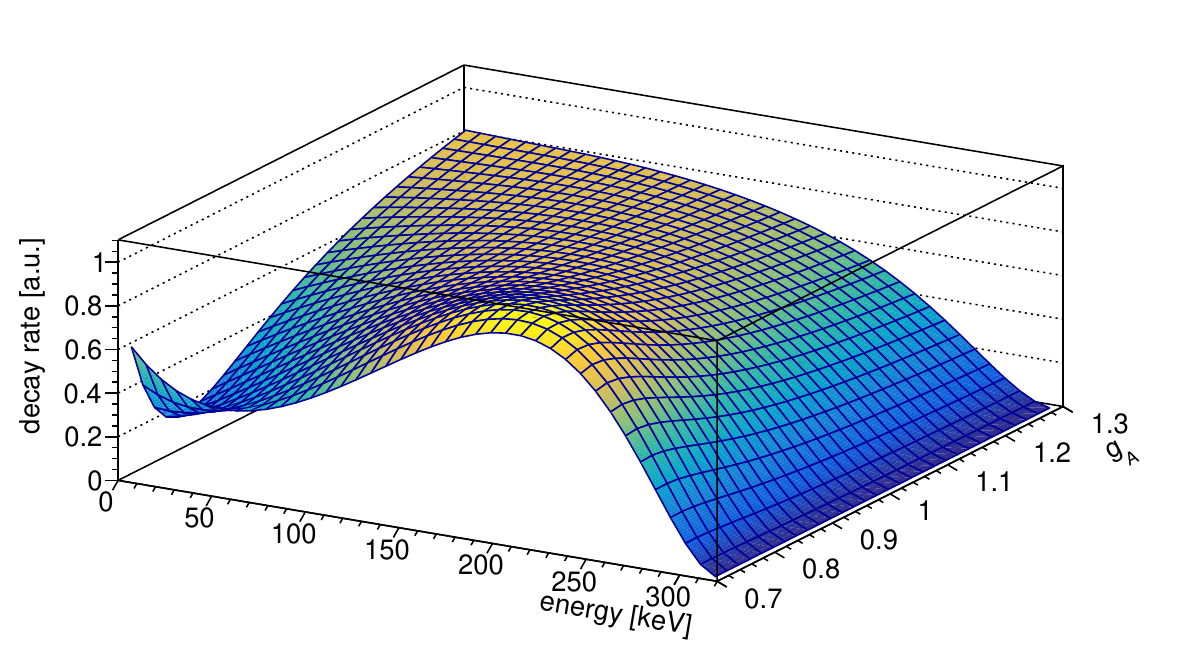}\\
\includegraphics[width=0.7\textwidth]{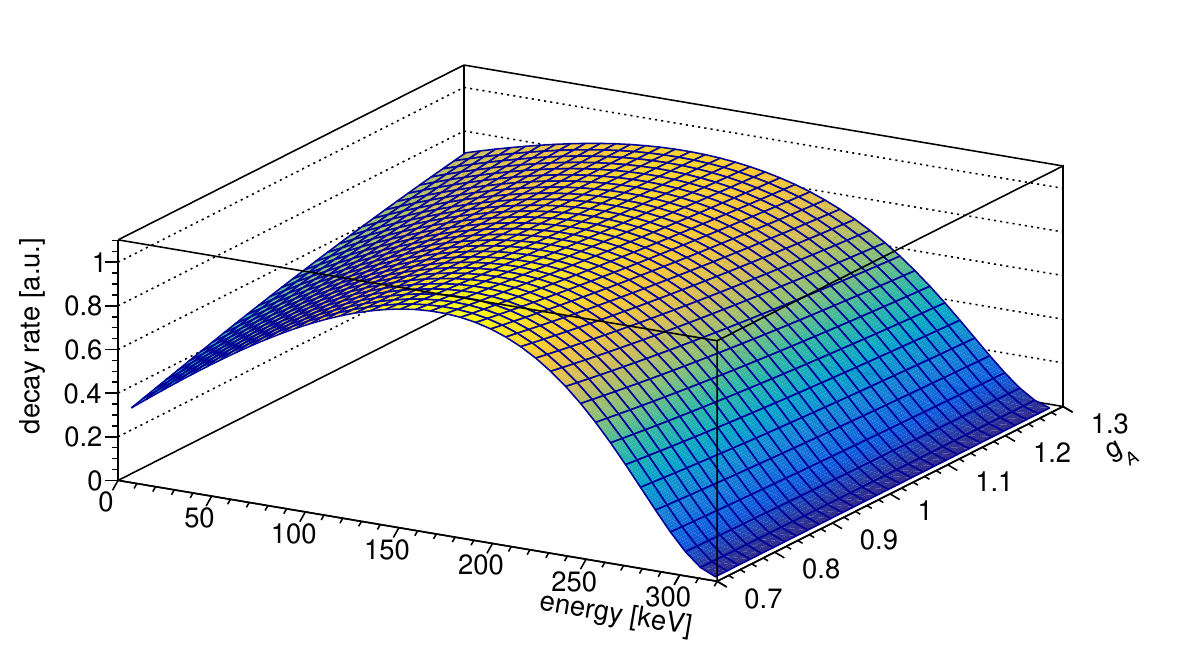}
\caption{Nuclear model predictions for the \cd113 spectrum-shape dependence on \gA under the revised SSM. \textit{Top:} ISM. \textit{Middle:} MQPM. \textit{Bottom:} IBFM-2.}
\label{fig:appendix_gA_dependence_cvc}
\end{figure*}

\section{Visualization of allowed spectrum-ranges}

\begin{figure*}[ht!]
\centering
\includegraphics[width=0.33\textwidth]{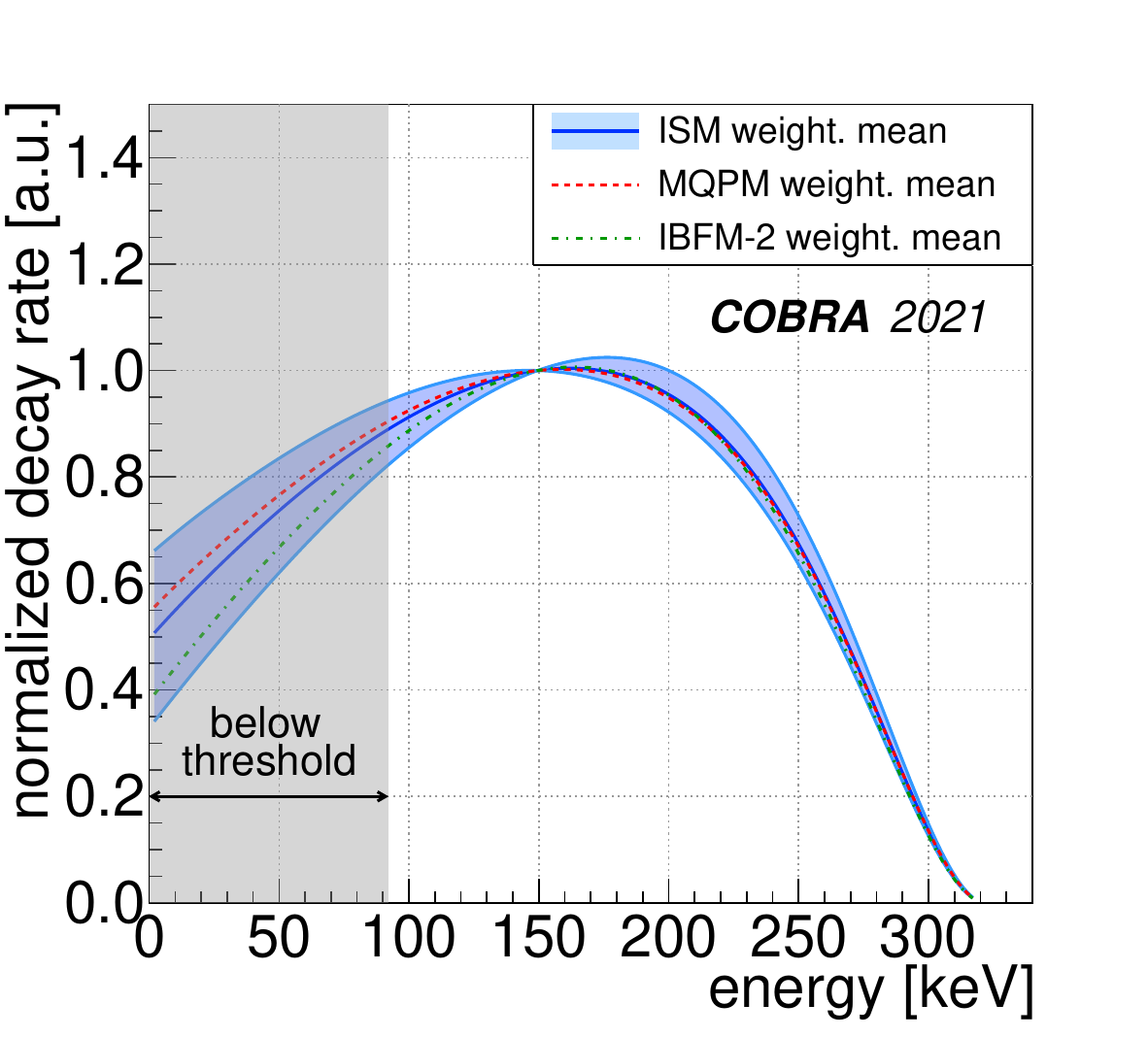}
\includegraphics[width=0.33\textwidth]{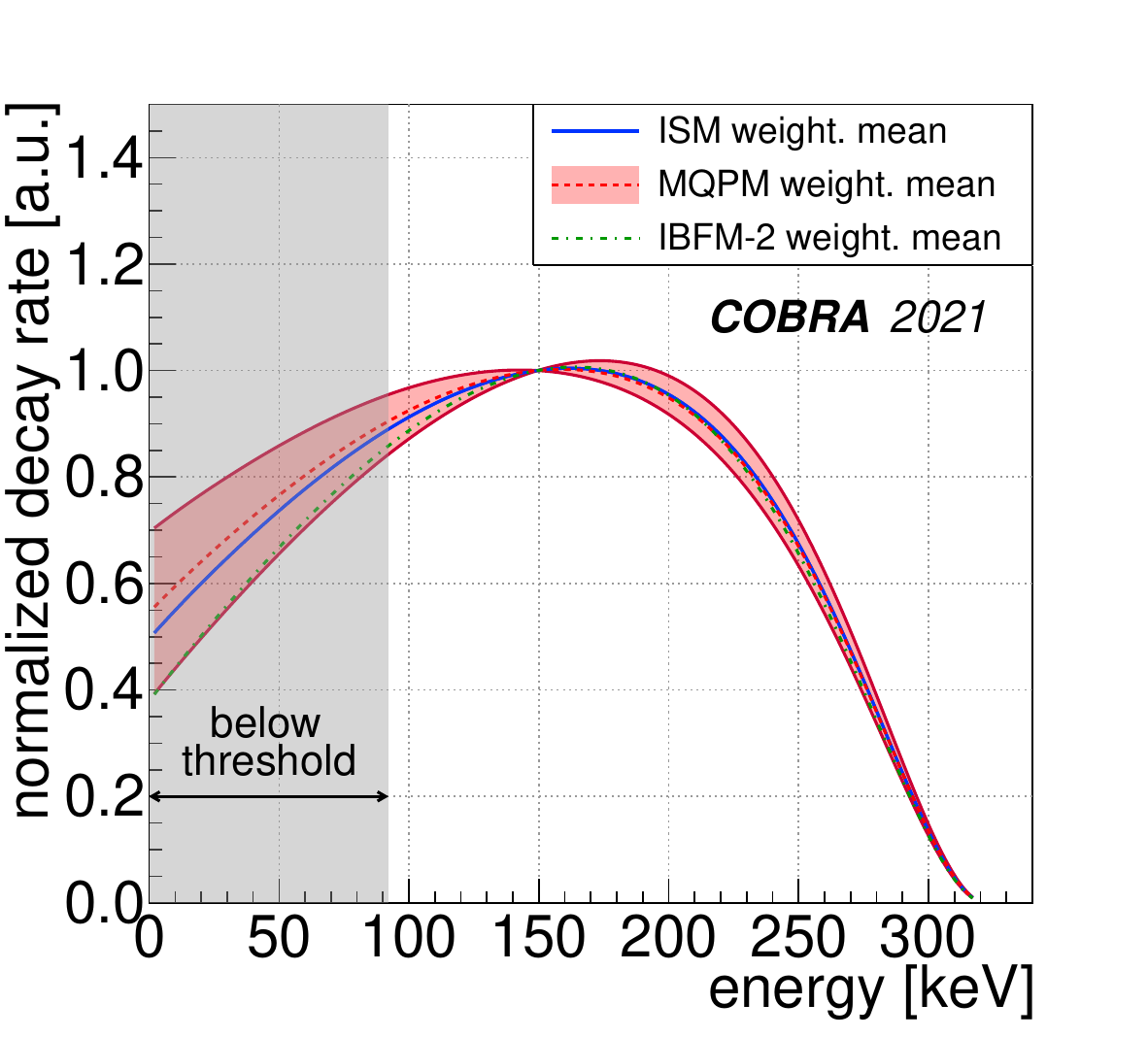}
\includegraphics[width=0.33\textwidth]{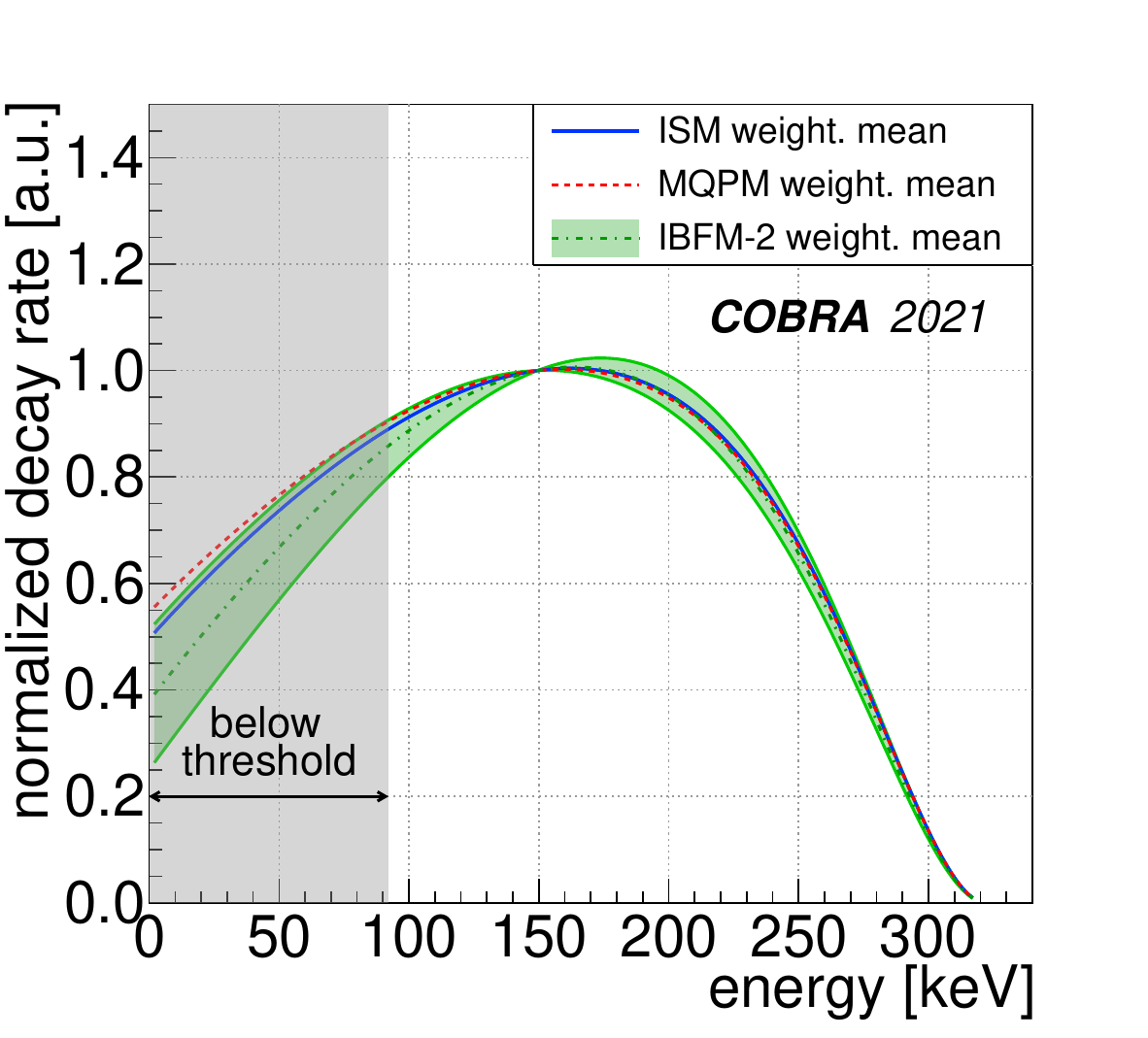}
\caption{Allowed spectrum range for the \cd113 $\beta$-decay according to the \gA range resulting from the spectrum-shape analysis. \textit{Left:} ISM. \textit{Middle:} MQPM. \textit{Right:} IBFM-2. A spline interpolation is used to construct template spectra for $\overline{g}_\text{A} \pm \overline{\sigma}_\text{sys}$ from the originally provided templates including the determined uncertainties for each model. For comparison the templates that correspond to the best match $\overline{g}_\text{A}$ results of the respective other two models are shown in each of the three graphics.}
\label{fig:Cd113_allowed_spectrum_range}
\end{figure*}

\newpage
\section{Visualization of single detector results}

\begin{figure*}[ht!]
\centering
\includegraphics[width=0.9\textwidth]{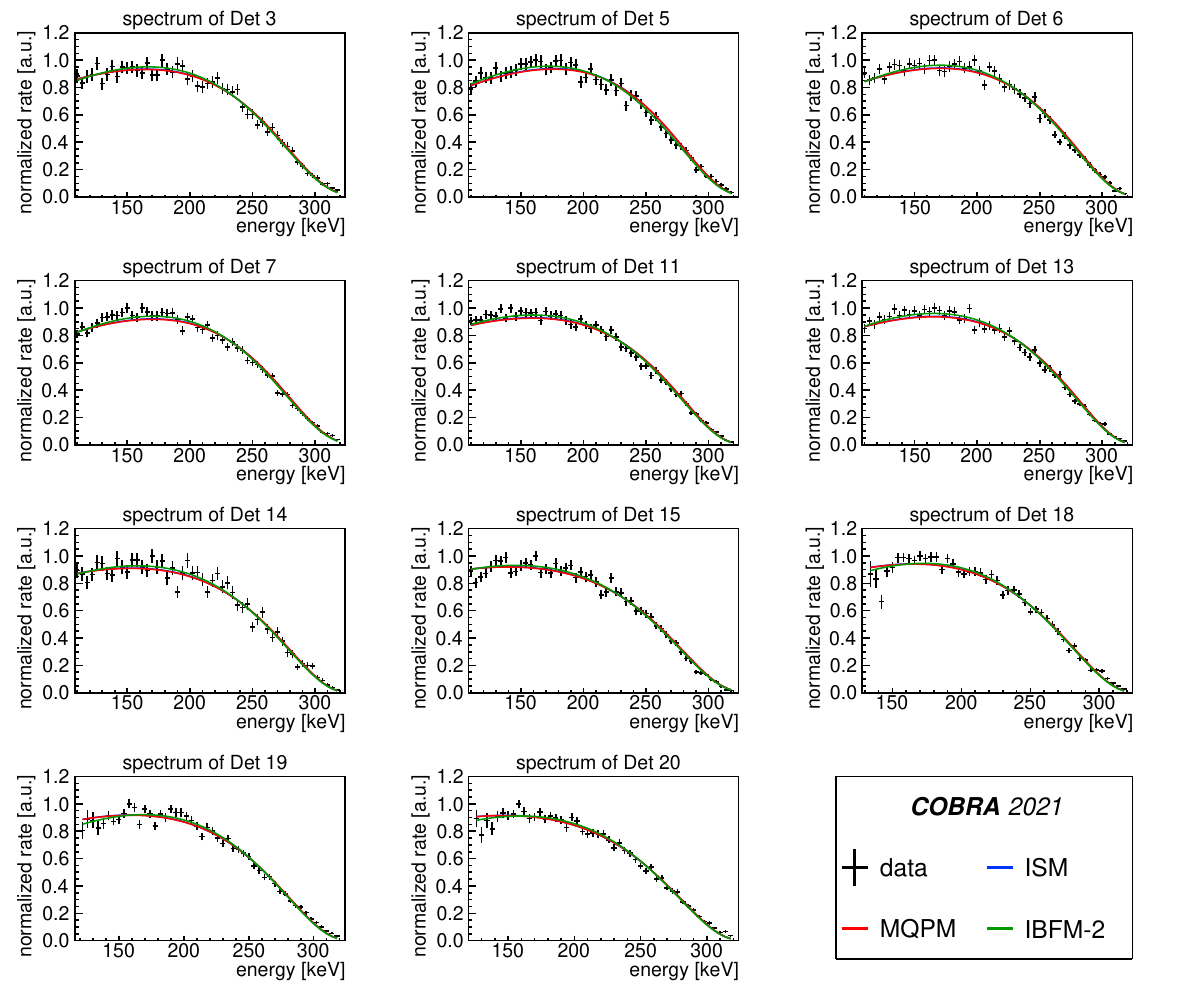}
\caption{Compilation of the single detector results (Det 3 -- Det 20). Each graphic displays the experimental data of a single COBRA detector and the respective best match templates for the ISM, MQPM and IBFM-2 according to the spectrum shape comparison.}
\label{fig:appendix_single_detector_results01}
\end{figure*}

\begin{figure*}[ht!]
\centering
\includegraphics[width=0.9\textwidth]{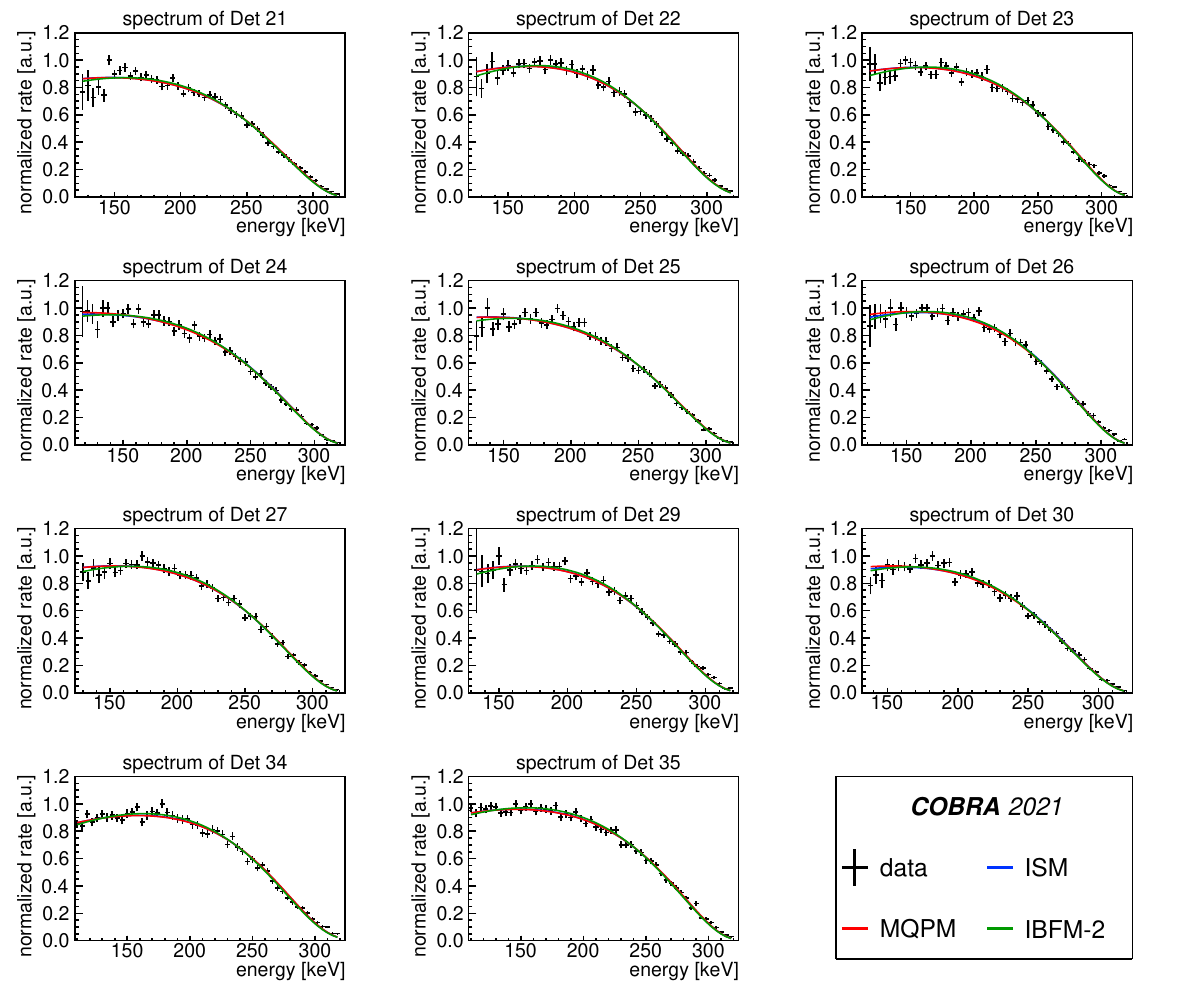}
\caption{Compilation of the single detector results (Det 21 -- Det 35). Each graphic displays the experimental data of a single COBRA detector and the respective best match templates for the ISM, MQPM and IBFM-2 according to the spectrum shape comparison.}
\label{fig:appendix_single_detector_results02}
\end{figure*}

\begin{figure*}[ht!]
\centering
\includegraphics[width=0.9\textwidth]{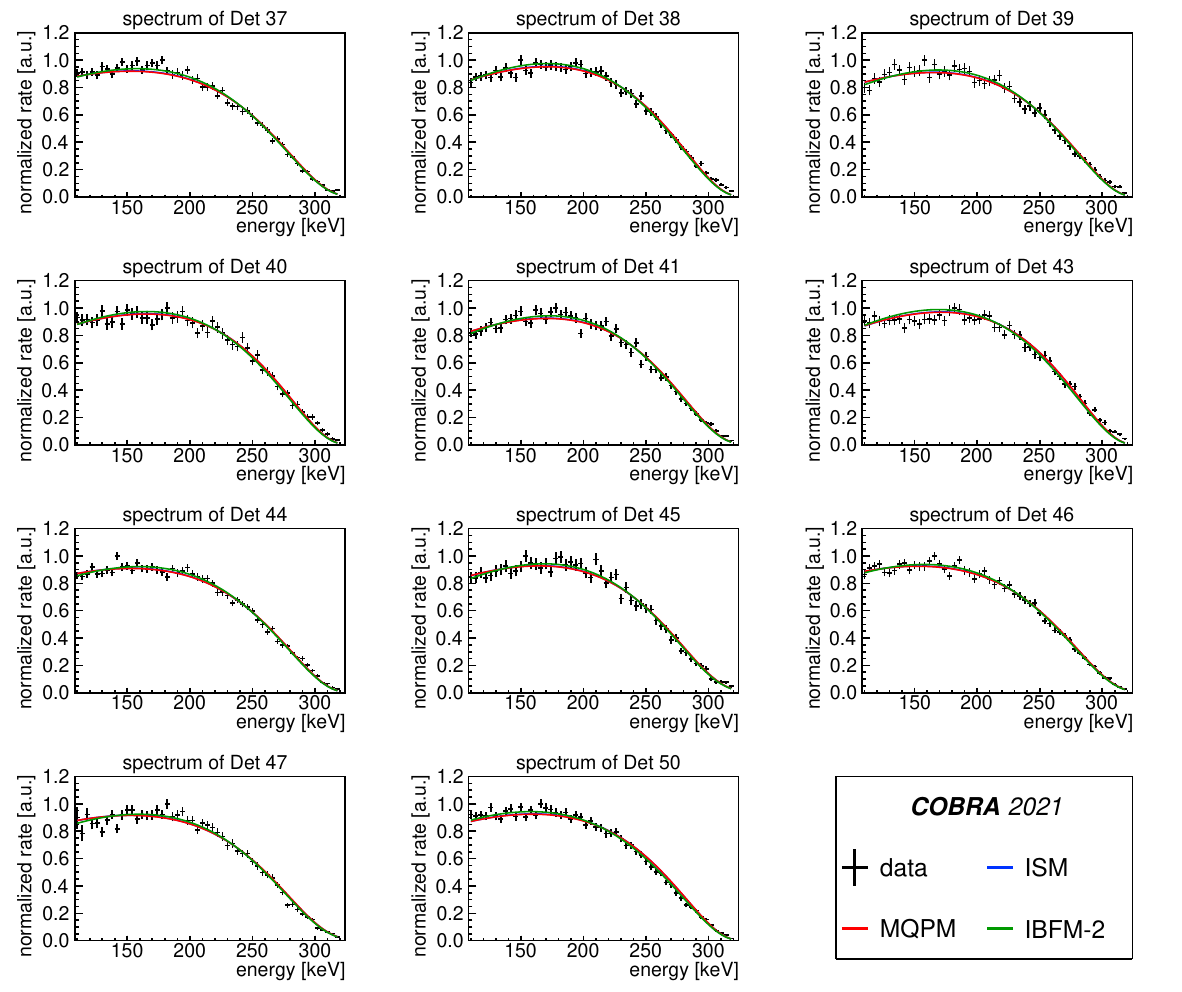}
\caption{Compilation of the single detector results (Det 37 -- Det 50). Each graphic displays the experimental data of a single COBRA detector and the respective best match templates for the ISM, MQPM and IBFM-2 according to the spectrum shape comparison.}
\label{fig:appendix_single_detector_results03}
\end{figure*}

\begin{figure*}[ht!]
\centering
\includegraphics[width=0.9\textwidth]{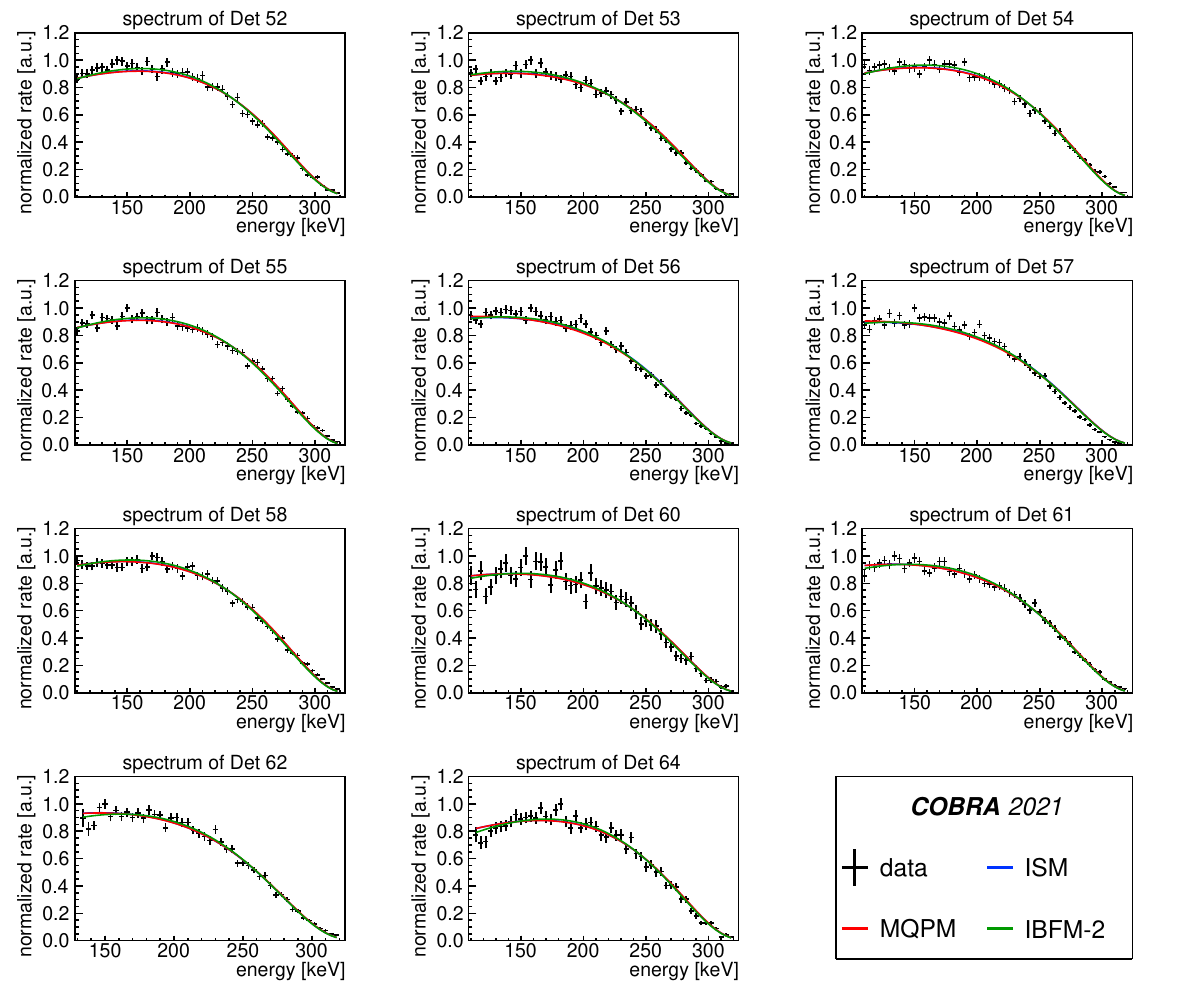}
\caption{Compilation of the single detector results (Det 52 -- Det 64). Each graphic displays the experimental data of a single COBRA detector and the respective best match templates for the ISM, MQPM and IBFM-2 according to the spectrum shape comparison.}
\label{fig:appendix_single_detector_results04}
\end{figure*}


%%       \citet{<label>} ==> Jones et al. [21]
%%       \citep{<label>} ==> [21]
%%



\end{document}